\documentclass[
 preprint, 
 amsmath,amssymb,
 aps, physrev,
]{revtex4-2}

\usepackage{graphicx}
\usepackage{dcolumn}
\usepackage{bm}

\begin{document}

\preprint{}

\title{\textbf{Inverse-Designed Lithium Niobate Wavelength Demultiplexer via Birefringent Effective Index Approximation}}

\author{Chihyeon Kim}
\thanks{These authors contributed equally to this work.}
\affiliation{%
Department of Electronic Engineering, Hanyang University, Seoul 04763, Republic of Korea
}%

\author{Minho Choi}
\thanks{These authors contributed equally to this work.}
\affiliation{%
Center for Quantum Technology, Korea Institute of Science and Technology (KIST), Seoul 02792, Republic of Korea
}%
\affiliation{%
Department of Artificial Intelligence Semiconductor Engineering, Hanyang University, Seoul 04763, Republic of Korea
}%

\author{Munseong Bae}
\affiliation{%
Department of Electronic Engineering, Hanyang University, Seoul 04763, Republic of Korea
}%

\author{Hyounghan Kwon}
\email{Corresponding author: hyounghankwon@kist.re.kr}
\affiliation{%
Center for Quantum Technology, Korea Institute of Science and Technology (KIST), Seoul 02792, Republic of Korea
}%
\affiliation{%
Division of Quantum Information, KIST School, Korea University of Science and Technology, Seoul 02792, Republic of Korea
}%

\author{Haejun Chung}
\email{Corresponding author: haejun@hanyang.ac.kr}
\affiliation{%
Department of Electronic Engineering, Hanyang University, Seoul 04763, Republic of Korea
}%
\affiliation{%
Department of Artificial Intelligence Semiconductor Engineering, Hanyang University, Seoul 04763, Republic of Korea
}%

\date{\today}

\begin{abstract}
Inverse design of thin-film lithium niobate (TFLN) photonic devices is computationally demanding because optical birefringence and fabrication-induced slanted sidewalls generally require three-dimensional electromagnetic models. We introduce a birefringent effective-index (BEI) method to reduce this problem to two dimensions while retaining polarization-dependent slab confinement and a representative cross section of the etched geometry. The method is integrated with adjoint topology optimization and fabrication constraints to design a $30~\mu\mathrm{m}\times10~\mu\mathrm{m}$ demultiplexer that routes 1550 and 775~nm light to separate output ports. Quantitative comparisons with three-dimensional finite-difference time-domain simulations establish the accuracy and etch-depth dependence of the reduced model. The fabricated device provides mean signal-to-crosstalk ratios of 13.9~dB across 1540--1560~nm and 13.3~dB across 770--780~nm. A two-stage cascaded configuration increases the output extinction ratio to 26.8~dB in the telecom band and 20.3~dB in the near-visible band. This fabrication-aware reduced-dimensional approach enables optimizations of the multifunctional photonic devices for nonlinear optical and quantum applications on the TFLN platform.

\begin{description}
\item[Keywords] {thin-film lithium niobate, inverse design, birefringence, wavelength demultiplexing, topology optimization}
\end{description}
\end{abstract}

\maketitle


\section{\label{sec:level1}Introduction}

\begin{figure*}[tp]
\centerline{\includegraphics[width=\linewidth]{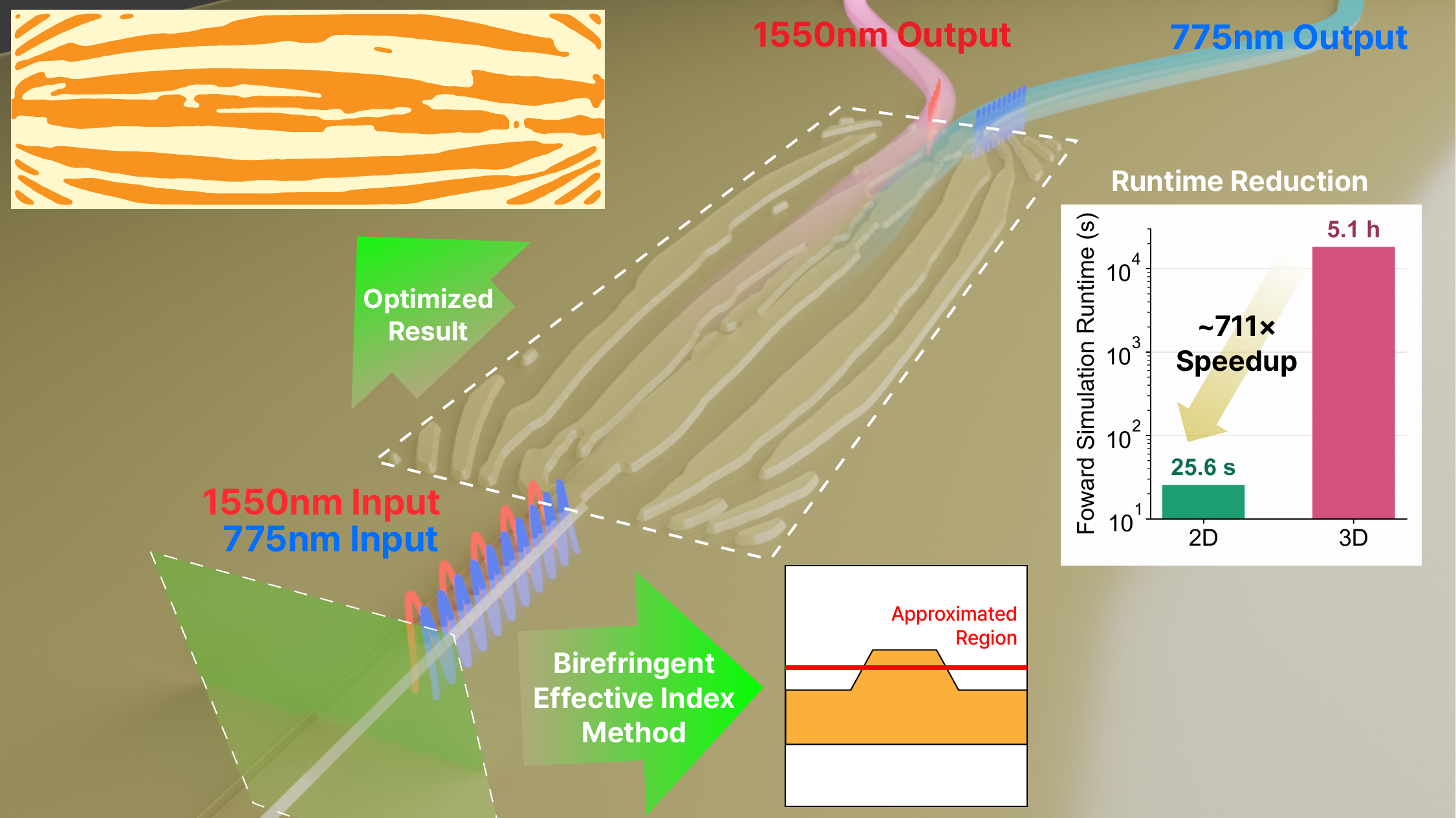}}
\caption{\fontsize{9pt}{7.5pt}\selectfont
Schematic illustration of the operating principle of the inverse-designed LN DEMUX proposed in this work. When light at wavelengths of 1550 nm and 775 nm is launched into the input port, the 1550 nm light is routed to the upper output port, whereas the 775 nm light is directed to the lower output port. For the two-dimensional optimization, we introduce a birefringent effective-index method based on the center of the slanted sidewall cross section. As shown in the inset on the right, this approach achieves a 711-fold reduction in computational time compared with full three-dimensional simulation. (identical hardware; see Methods)\label{fig1}}
\end{figure*}
Inverse design has become a powerful approach for developing compact nanophotonic devices through performance-driven and fabrication-aware optimization~\cite{molesky_inverse_2018, han_breaking_2025, seo_physics-guided_2026, hammond_photonic_2021}. Among these methods, adjoint optimization is particularly efficient because the gradient of a figure of merit with respect to a large number of design variables can be obtained using one forward and one adjoint simulation for a given objective~\cite{miller_photonic_2012}. However, achieving the desired performance typically requires hundreds of optimization iterations, and complex photonic structures often require three-dimensional full-wave simulations~\cite{shang_inverse-designed_2023, he_broadband_2025, lyu_inverse-designed_2026, lyu_inverse_2023, kim_suppressing_2026}. Repeated three-dimensional simulations therefore impose a substantial computational cost, which limits efficient exploration of large design spaces.

To reduce this computational cost, recent studies have employed two-dimensional effective-index methods (EIMs), which approximate vertical optical confinement using effective refractive indices derived from slab modes~\cite{hammer_effective_2009, nikkhah_inverse-designed_2024, qiao_fabrication-aware_2025}. Such reduced-dimensional models can provide accurate and efficient optimization for structures with limited vertical variation, particularly in isotropic materials and shallow-etch geometries~\cite{nikkhah_inverse-designed_2024}. For materials with in-plane optical anisotropy, however, conventional scalar EIMs cannot fully represent the polarization- and direction-dependent optical response. Inverse design of these structures has therefore largely relied on full three-dimensional simulations~\cite{shang_inverse-designed_2023, he_broadband_2025, lyu_inverse-designed_2026, lyu_inverse_2023, kim_suppressing_2026}.

Optically anisotropic materials, including lithium niobate (LN)~\cite{zhu_integrated_2021}, aluminum nitride~\cite{majkic_optical_2015,liu_aluminum_2023}, barium titanate~\cite{karvounis_barium_2020}, and emerging two-dimensional materials~\cite{ermolaev_giant_2021,li_review_2021}, have attracted considerable interest for integrated photonics because they offer nonlinear, electro-optic, acousto-optic, and other active optical functionalities. Their anisotropic optical response, however, complicates reduced-dimensional modeling and often necessitates computationally demanding three-dimensional simulations.

Among these materials, LN is particularly attractive for nonlinear and quantum photonic devices based on second-harmonic generation (SHG)~\cite{wang_ultrahigh-efficiency_2018, lu_periodically_2019} and spontaneous parametric down-conversion (SPDC)~\cite{zhao_high_2020}, as well as electro-optic~\cite{wang_integrated_2018, zhang_broadband_2019} and acousto-optic devices~\cite{xu_unveiling_2024, cai_acousto-optical_2019}. In many nonlinear LN systems, SHG and SPDC connect optical fields near 1550 and 775~nm, corresponding to an octave-spanning wavelength range~\cite{boyd_nonlinear_2008}. Efficient separation of the pump, harmonic, signal, or idler fields is therefore important for integrated nonlinear circuits and quantum light sources~\cite{zhao_high_2020, lu_periodically_2019, guo_parametric_2017}. Such wavelength separation has often relied on external filtering, motivating recent efforts toward on-chip implementations~\cite{nehra_few-cycle_2022,ledezma_octave-spanning_2023,dean_low-power_2026}. Integrated demultiplexers based on resonators~\cite{rabus_integrated_2020,dahlem_reconfigurable_2011,chan_ultra-wide_2022}, Bragg gratings~\cite{zhang_bragg_2025,zhu_hybrid_2025,liu_ultra-compact_2023}, and interferometric structures~\cite{yen_fabrication-tolerant_2021,zhang_full_2018,jeong_experimental_2024,gorgulu_ultra-broadband_2023} can require extended interaction lengths, cascaded sections, or wavelength-sensitive resonances. Compact inverse-designed demultiplexers provide an alternative, although their optimization in anisotropic LN remains limited by the high computational cost of three-dimensional simulations.

Conventional two-dimensional EIMs approximate vertical optical confinement using scalar effective indices and are most accurate for isotropic structures with limited vertical variation~\cite{hammer_effective_2009}. Their agreement with full three-dimensional simulations generally decreases as the etch depth increases, which often restricts their use to shallowly etched geometries~\cite{nikkhah_inverse-designed_2024}. This approximation becomes more challenging in X-cut LN because the optical axis lies in the film plane, causing the effective refractive index to depend on both propagation direction and polarization~\cite{han_breaking_2025, yi_anisotropy-free_2024}. A single scalar effective index therefore cannot adequately describe a freeform structure containing multiple local propagation directions.

Related 2.5D methods such as varFDTD in Ansys Lumerical MODE~\cite{noauthor_mode_nodate} represent vertical confinement through a scalar effective-index map derived from a selected reference slab mode~\cite{hammer_effective_2009}. Such a scalar representation does not capture the direction-dependent optical response of an in-plane anisotropic material. The modeling difficulty is further increased by the slanted sidewalls produced during LN dry etching~\cite{shang_inverse-designed_2023,kwon_photon-pair_2024,han_breaking_2025,lyu_inverse_2023,mu_polarization-insensitive_2026,qiao_fabrication-aware_2025}. Previous studies have therefore addressed these limitations by using crystal orientations that are isotropic in the film plane~\cite{qiao_fabrication-aware_2025}, performing full three-dimensional optimization~\cite{lyu_inverse-designed_2026,lyu_inverse_2023,he_broadband_2025}, or avoiding direct patterning of the LN layer~\cite{han_breaking_2025}.

In prior TFLN inverse-design studies, two-dimensional EIM has primarily been applied to z-cut LN, where the film plane is optically isotropic~\cite{qiao_fabrication-aware_2025}. This orientation, however, is less suitable for applications that seek to exploit the largest nonlinear and electro-optic coefficients of LN. Accessing $d_{33}=-27~\mathrm{pm/V}$ and $r_{33}=30~\mathrm{pm/V}$ requires strong electric-field components along the extraordinary axis. In X-cut LN, this axis lies in the film plane, making this crystal orientation widely used for nonlinear and electro-optic TFLN devices~\cite{zhu_integrated_2021}.

Inverse design of X-cut TFLN devices has therefore largely relied on full three-dimensional optimization~\cite{lyu_inverse-designed_2026,lyu_inverse_2023}. Even when varFDTD is used for preliminary optimization, three-dimensional simulations have remained necessary for subsequent design refinement or verification~\cite{he_broadband_2025}. Another approach avoids direct LN patterning by introducing an isotropic SiN layer on unetched LN~\cite{han_breaking_2025}, at the cost of additional material processing and reduced optical confinement in LN. These limitations motivate a reduced-dimensional model that retains the in-plane anisotropy of X-cut LN while approaching the computational cost of two-dimensional simulation.

To address these limitations, we introduce a birefringent effective-index (BEI) method and apply it to the inverse design of a compact 1550/775~nm wavelength demultiplexer with a $30\times10~\mu\mathrm{m}^2$ footprint on X-cut TFLN. The method derives effective indices for the ordinary and extraordinary polarization directions and incorporates them into an anisotropic permittivity tensor for two-dimensional simulation. This treatment retains the direction-dependent birefringent response that is absent from conventional scalar EIMs. Slanted sidewalls are approximated using the mid-plane cross section of the etched geometry, while fabrication-aware minimum linewidth and spacing constraints account for the corresponding width variation. Under identical computational resources, the resulting two-dimensional model reduces the forward-simulation time by approximately 711-fold compared with three-dimensional simulation, enabling efficient exploration of design parameters including etch depth. The method and optimized device are summarized in Fig.~\ref{fig1}. The final designs are evaluated using full three-dimensional simulations and experimentally demonstrated after fabrication. To the best of our knowledge, this is the first inverse design of an X-cut TFLN device completed without a three-dimensional optimization stage. The BEI method can also be extended to other birefringent materials, including lithium tantalate and barium titanate, by incorporating their corresponding polarization-dependent effective indices.

\section{Simulation Results}

\subsection{Birefringent Effective Index (BEI) Method}

The BEI method extends the conventional effective-index approximation to account for the in-plane birefringence of X-cut TFLN and the slanted geometry produced by partial etching. This subsection first summarizes the dimensional reduction used in conventional EIMs and then describes how polarization-dependent effective indices and a representative etched cross section are incorporated into the two-dimensional model. This reduced model is subsequently used for inverse design and evaluated against three-dimensional simulations.

EIMs reduce computational cost by approximating a three-dimensional waveguide structure with a two-dimensional model~\cite{hammer_effective_2009,b_inverse_2022,radford_inverse_2025,nikkhah_inverse-designed_2024}. The vertical cross section of each material region is treated as a slab waveguide, and the corresponding guided-mode effective index is assigned to that region in the two-dimensional plane. This approximation captures the dominant effect of vertical optical confinement while substantially reducing the number of simulation cells. EIMs have therefore been widely used for efficient simulation and inverse design of integrated photonic devices~\cite{b_inverse_2022,radford_inverse_2025}. They have also been applied to partially etched structures by assigning different effective indices to etched and unetched slab regions. For example, such a reduced-dimensional treatment was used to design optical matrix-multiplication devices that were subsequently fabricated and experimentally characterized~\cite{nikkhah_inverse-designed_2024}.

Most conventional EIMs were developed for isotropic materials such as silicon, where each region can be represented by a scalar effective index~\cite{b_inverse_2022,nikkhah_inverse-designed_2024}. This scalar treatment is inadequate for LN because its optical response depends on crystal orientation and field polarization. In addition, LN dry etching typically produces slanted sidewalls, causing the lateral waveguide geometry to vary through the film thickness~\cite{shang_inverse-designed_2023,kwon_photon-pair_2024,qiao_fabrication-aware_2025}. To address these effects, we introduce the BEI method, which assigns polarization-dependent effective indices derived from LN slab modes and represents the slanted geometry using a selected cross section of the etched structure.

\begin{figure*}[tp]
\centerline{\includegraphics[width=\linewidth]{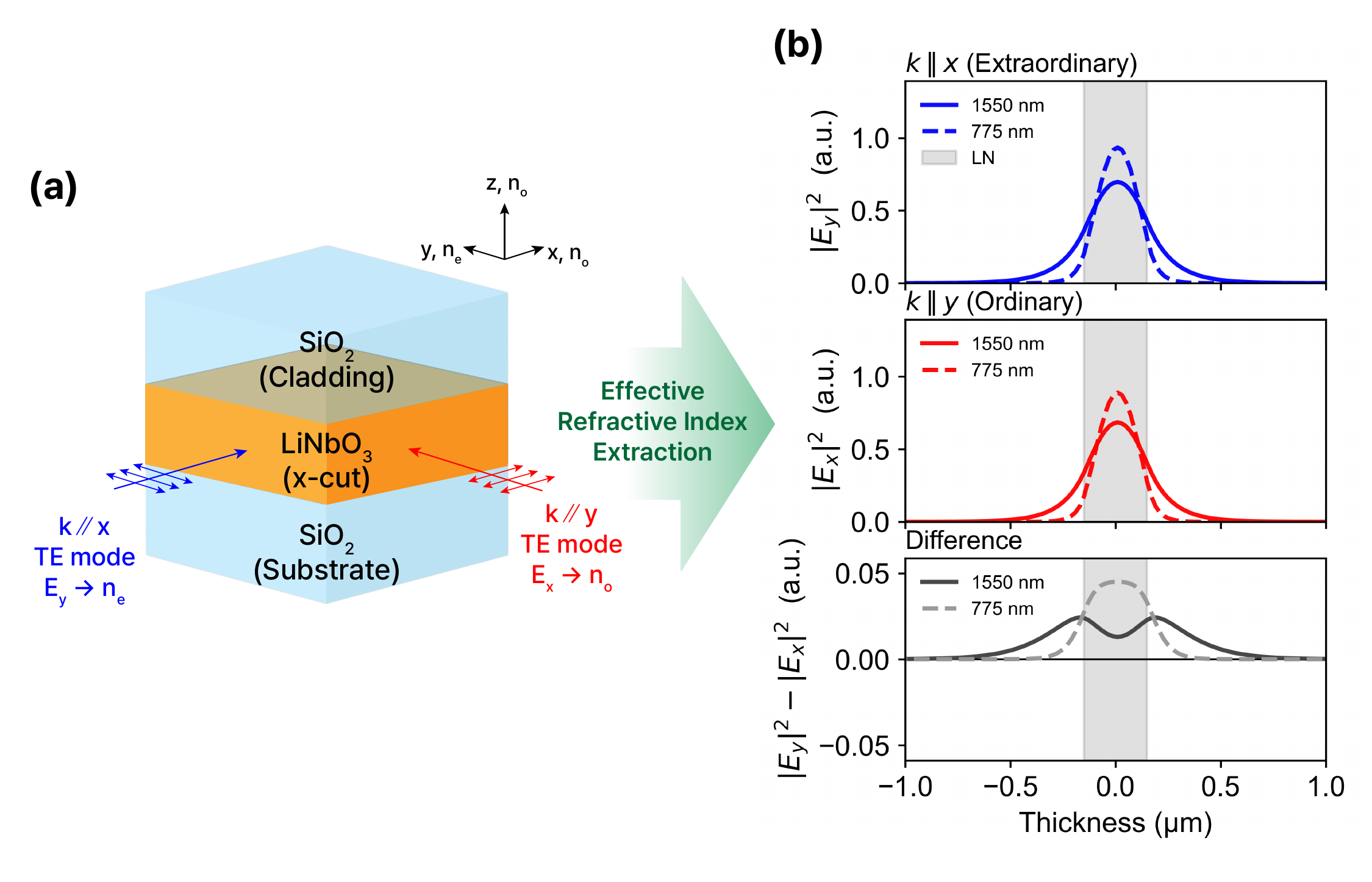}}
\caption{\fontsize{9pt}{7.5pt}\selectfont
Schematic illustration of birefringent effective index extraction used to determine the effective index of a birefringent medium according to its crystallographic orientation using planar slab waveguides. (a) Slab waveguide composed of a SiO$_2$ substrate, a LiNbO$_3$ core, and a SiO$_2$ cladding, where the LN layer has an X-cut crystal orientation. The dominant electric-field component of the fundamental TE mode for each propagation direction and the corresponding refractive-index component that primarily governs the mode are illustrated. (b) Calculated electric-field distributions for the slab waveguides. The dominant field components of the TE modes are plotted for wavelengths of 1550 nm (solid lines) and 775 nm (dashed lines). The upper panel shows the intensity of the $E_y$ component for propagation along the $x$ direction (blue), while the middle panel shows the intensity of the $E_x$ component for propagation along the $y$ direction (red). The lower panel presents the difference between the dominant electric-field components in gray, revealing that the field distributions vary because of the birefringence of LN.\label{fig2}}
\end{figure*}

As shown in Fig.~\ref{fig2}(a), the TE-like mode propagating along the $x$-direction is dominated by $E_y$ and therefore primarily samples the extraordinary permittivity of X-cut LN. For propagation along the $y$-direction, the dominant $E_x$ component primarily samples the ordinary permittivity. Figure~\ref{fig2}(b) shows the corresponding field distributions in a 300-nm-thick LN slab at 775 and 1550~nm. Because the extraordinary refractive index is lower than the ordinary refractive index at both wavelengths, the mode associated with the extraordinary response exhibits a lower effective index and weaker vertical confinement. Field distributions for other residual LN thicknesses are provided in Fig.~S1, and further details of the BEI formulation are given in the Methods section.

\begin{table*}[!ht]
\caption{\label{tab:effective_index}
Effective indices used for the birefringent effective-index model.
}
\begin{ruledtabular}
\begin{tabular*}{\textwidth}{@{\extracolsep{\fill}}lccccc}
\textrm{Etch depth} &
\textrm{40 nm} &
\textrm{60 nm} &
\textrm{80 nm} &
\textrm{100 nm} &
\textrm{120 nm} \\
\colrule
$n_{\mathrm{subs},e,1550}$ & 1.7373 & 1.7122 & 1.6874 & 1.6595 & 1.6322 \\
$n_{\mathrm{subs},e,775}$  & 1.9759 & 1.9528 & 1.9311 & 1.9005 & 1.8713 \\
$n_{\mathrm{subs},o,1550}$ & 1.7865 & 1.7586 & 1.7310 & 1.6995 & 1.6685 \\
$n_{\mathrm{subs},o,775}$  & 2.0537 & 2.0293 & 2.0067 & 1.9739 & 1.9428 \\
\colrule
$n_{\mathrm{str},e,1550}$ & \multicolumn{5}{c}{${1.781}^{\dagger}$} \\
$n_{\mathrm{str},e,775}$  & \multicolumn{5}{c}{2.0099} \\
$n_{\mathrm{str},o,1550}$ & \multicolumn{5}{c}{1.8347} \\
$n_{\mathrm{str},o,775}$  & \multicolumn{5}{c}{2.0890} \\
\end{tabular*}
\end{ruledtabular}

\vspace{2pt}
\begin{minipage}{\textwidth}
\footnotesize
$n_{\mathrm{subs}}$ and $n_{\mathrm{str}}$ denote the effective indices of the residual slab and patterned structure, respectively.\\
The subscripts $e$ and $o$ indicate the extraordinary and ordinary directions, respectively.\\
$\dagger$ The structure effective indices correspond to the unetched region of the 300 nm-thick x-cut LN layer and are therefore independent of etch depth.
\end{minipage}
\end{table*}

Table~\ref{tab:effective_index} summarizes the slab-mode effective indices $n^{\mathrm{eff}}{i,j,k}$ for the etched and unetched regions, where $i$, $j$, and $k$ denote the etch condition, polarization direction, and wavelength, respectively. The polarization index $j$ represents the ordinary ($o$) or extraordinary ($e$) direction. Using these values, the effective-index tensor for each region was defined as $\mathbf{n}^{\mathrm{eff}}_{i,k}=\mathrm{diag}\left[n^{\mathrm{eff}}_{i,o,k},n^{\mathrm{eff}}_{i,e,k},n^{\mathrm{eff}}_{i,o,k}\right]$, corresponding to the bulk refractive-index tensor $\mathrm{diag}(n_o,n_e,n_o)$ of X-cut LN. These tensors were assigned to the corresponding regions of the two-dimensional model to retain the polarization-dependent birefringent response.

A further challenge in applying EIM to LN is the slanted sidewall produced by dry etching. For the resulting trapezoidal geometry, the accuracy of the BEI approximation depends on the cross section used to represent the three-dimensional structure. In this work, the waveguide width at the mid-plane of the etched LN region was selected as the representative cross section. The top and bottom widths represent the extrema of the lateral variation and can therefore under- or overestimate the modal effective index. The mid-plane width provides an intermediate geometry that reduces this bias within a single-cross-section approximation. Although this representation does not reproduce the full slanted profile, it provides a simple and physically reasonable approximation for the BEI model.

To evaluate this approximation, we compared the modal effective indices obtained from the one-dimensional slab model, a two-dimensional rectangular-cross-section MPB model, and a two-dimensional slanted-cross-section MPB model for each propagation direction and wavelength band, as shown in Figs.~S2--S5. Over waveguide widths of 0.5--2~\textmu m, the effective indices predicted by the slab-based BEI approximation agree with those of the slanted-cross-section reference model with a mean deviation below $\Delta n_{\mathrm{eff}}\approx0.004$ and a maximum deviation below 0.01. The deviation increases for narrower waveguides, where a larger fraction of the modal field interacts with the slanted sidewall region. The overall agreement supports the use of the mid-plane cross section as a representative geometry in the BEI model. Further details are provided in Supporting Information Section~S2.

\subsection{BEI-based Optimization Results}

\begin{figure*}[t!]
\centerline{    \includegraphics[
        width=\textwidth
    ]{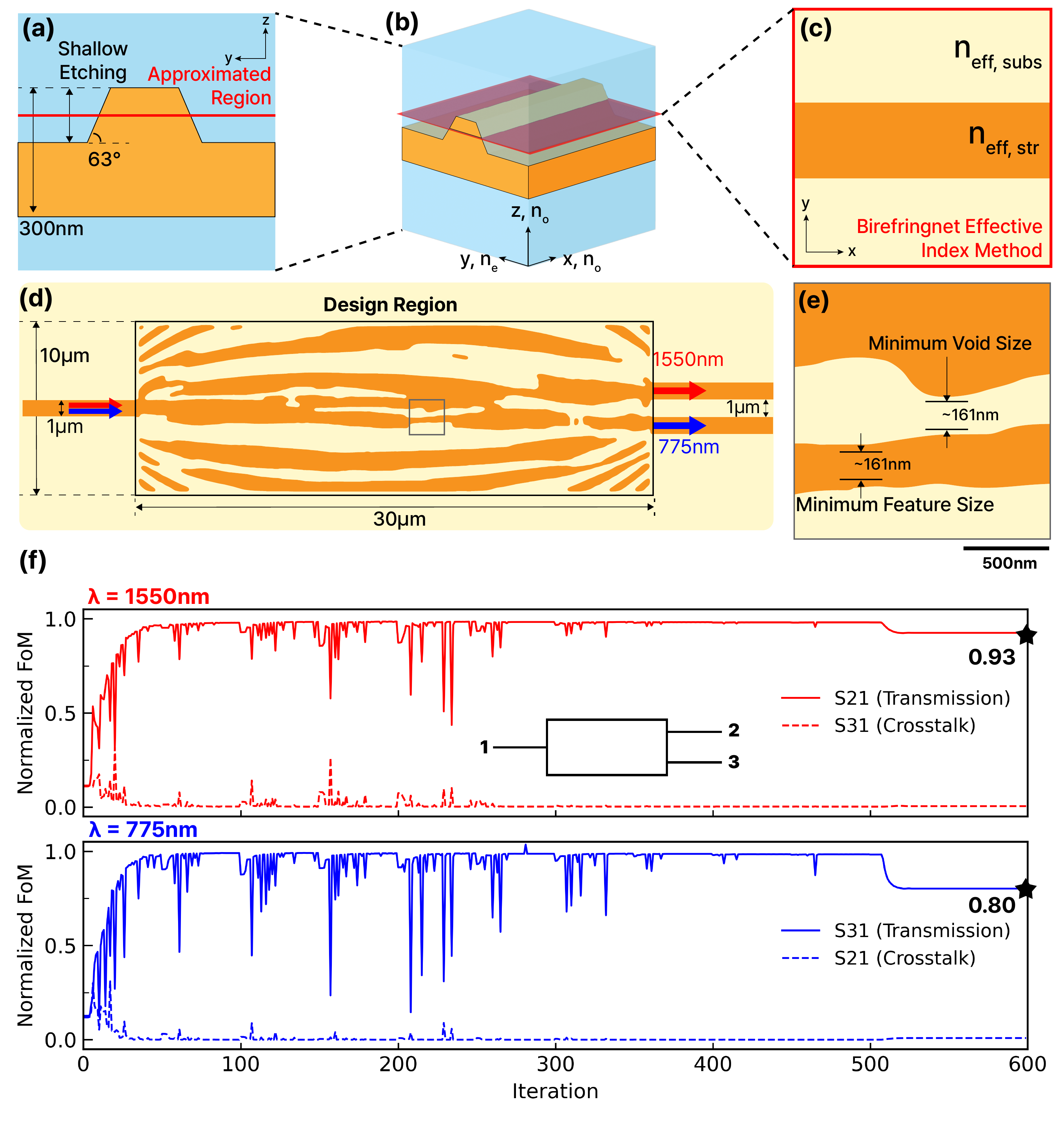}}
\caption{\fontsize{9pt}{7.5pt}\selectfont
Schematic illustration of the BEI method and the resulting optimized device. (a) Cross-sectional geometry of the 300 nm-thick LN layer, partially etched to form slanted sidewalls. The approximated region to which the BEI method is applied, indicated by the red line. (b) Three-dimensional schematic of an X-cut LN waveguide fabricated in this work and oriented along the ordinary propagation direction. The plane to which the BEI method is applied is indicated in red. (c) Waveguide structure represented in two dimensions using the BEI method. The dark-orange region corresponds to the structure effective index, while the light-yellow region corresponds to the substrate effective index. (d) Overall device configuration and geometry of the 1550 nm–775 nm demultiplexer obtained through adjoint optimization. (e) Visualization confirming that the optimized geometry satisfies the minimum feature-size and minimum void-size constraints. (f) Evolution of the normalized figure of merit (FoM) according to optimization iteration for the 1550 nm band (top) and the 775 nm band (bottom). The final optimized structure shown in the figure corresponds to iteration 600, marked by a star.\label{fig3}}
\end{figure*}

Figure~\ref{fig3} illustrates the BEI representation and its application to the wavelength demultiplexer. The partially etched 300-nm-thick LN layer has slanted sidewalls with an angle of approximately $63^\circ$, as shown in Fig.~\ref{fig3}(a). The ridge width at the mid-plane of the etched LN region was selected to represent the trapezoidal cross section in the reduced-dimensional model. Figure~\ref{fig3}(b) shows the corresponding three-dimensional X-cut LN waveguide, while Fig.~\ref{fig3}(c) shows its two-dimensional representation. The etched region was assigned the effective-index tensor of the residual slab,
$\mathbf{n}^{\mathrm{eff}}_{\mathrm{sub},k}=\mathrm{diag}\!\left[n^{\mathrm{eff}}_{\mathrm{sub},o,k},n^{\mathrm{eff}}_{\mathrm{sub},e,k},n^{\mathrm{eff}}_{\mathrm{sub},o,k}\right]$,
whereas the unetched ridge was assigned
$\mathbf{n}^{\mathrm{eff}}_{\mathrm{str},k}=\mathrm{diag}\!\left[n^{\mathrm{eff}}_{\mathrm{str},o,k},n^{\mathrm{eff}}_{\mathrm{str},e,k},n^{\mathrm{eff}}_{\mathrm{str},o,k}\right]$.
This assignment retains the polarization-dependent birefringent response of X-cut LN within the two-dimensional model.

The inverse-design geometry is shown in Fig.~\ref{fig3}(d). Light from a $1~\mu\mathrm{m}$-wide input waveguide enters a $30\times10~\mu\mathrm{m}^2$ design region and is routed to two output waveguides separated by a $1~\mu\mathrm{m}$ gap. The optimization directs the 1550-nm band to the upper output port, corresponding to $S_{21}$, and the 775-nm band to the lower output port, corresponding to $S_{31}$, while suppressing coupling to the undesired port. The design-region dimensions were selected from a size-dependent optimization study, which showed only marginal performance improvement for larger regions, as summarized in Fig.~S6.

Rather than applying an explicit wavelength correction for the spectral shift between the reduced-dimensional and three-dimensional models, the device was optimized over broad wavelength ranges of 1500--1600~nm and 750--800~nm to maintain routing performance under moderate spectral shifts. Five wavelengths were sampled in each band, with target transmission and crosstalk evaluated at both output ports, resulting in 20 competing objectives. These objectives were combined using a minimax formulation that improves the worst-performing condition during optimization~\cite{sigmund_topology_2013}. A conic density filter and smooth projection were applied throughout the optimization to suppress subresolution features and promote a binary material distribution~\cite{hammond_photonic_2021}. In the final stage, minimum feature- and void-size constraints corresponding to 161 nm were introduced. Owing to discretization on the simulation grid, the implemented constraint value was 160 nm.. The resulting binary structure is shown in Fig.~\ref{fig3}(e). Further details of the optimization formulation, filtering, geometric constraints, and optimizer settings are provided in the Methods section.

Figure~\ref{fig3}(f) shows the evolution of the normalized figures of merit over 600 optimization iterations. The periodic decreases in FoM every 50 iterations result from doubling the projection factor $\beta$, which sharpens the material distribution and temporarily reduces performance before subsequent optimization recovers it. These decreases become smaller as the design approaches a binary distribution and further increases in $\beta$ produce progressively smaller geometric changes. The larger drop at iteration 500 occurs when the minimum feature-size and minimum void-size constraints are activated, modifying subresolution features in the nearly converged design. At the final iteration, the normalized FoM reaches approximately 0.93 for the 1550-nm band and 0.80 for the 775-nm band while maintaining low crosstalk. The optimized two-dimensional geometry was then converted into a three-dimensional structure with slanted sidewalls and evaluated using full three-dimensional simulations. This comparison assesses the performance difference introduced by the BEI approximation and the reconstruction of the three-dimensional etched geometry.

\subsection{Three-Dimensional FDTD Verification}

Full three-dimensional FDTD simulations were performed to assess the accuracy of the BEI approximation across different etch depths. A 300-nm-thick X-cut LN film was placed between the SiO$_2$ substrate and upper cladding, and the computational domain was terminated by 1-$\mu\mathrm{m}$-thick perfectly matched layers along the $z$-direction.

To reconstruct the slanted sidewalls, the BEI-based optimized two-dimensional pattern was defined as the lateral cross section at the mid-plane of the etched LN region. Above this plane, the pattern was progressively eroded to decrease the ridge width toward the top surface, whereas below it, the pattern was progressively dilated to increase the ridge width toward the bottom surface. The lateral offset at each height was determined from the prescribed sidewall angle and the distance from the mid-plane. The resulting cross sections were generated using harmonic morphological operations~\cite{svanberg_density_2013} and stacked along the $z$-direction. The LN layer below the etched region remained unpatterned, producing a partially etched three-dimensional structure with trapezoidal features and a sidewall angle of $63^\circ$.

\begin{figure*}[tp]
\centerline{    \includegraphics[
        width=\textwidth
    ]{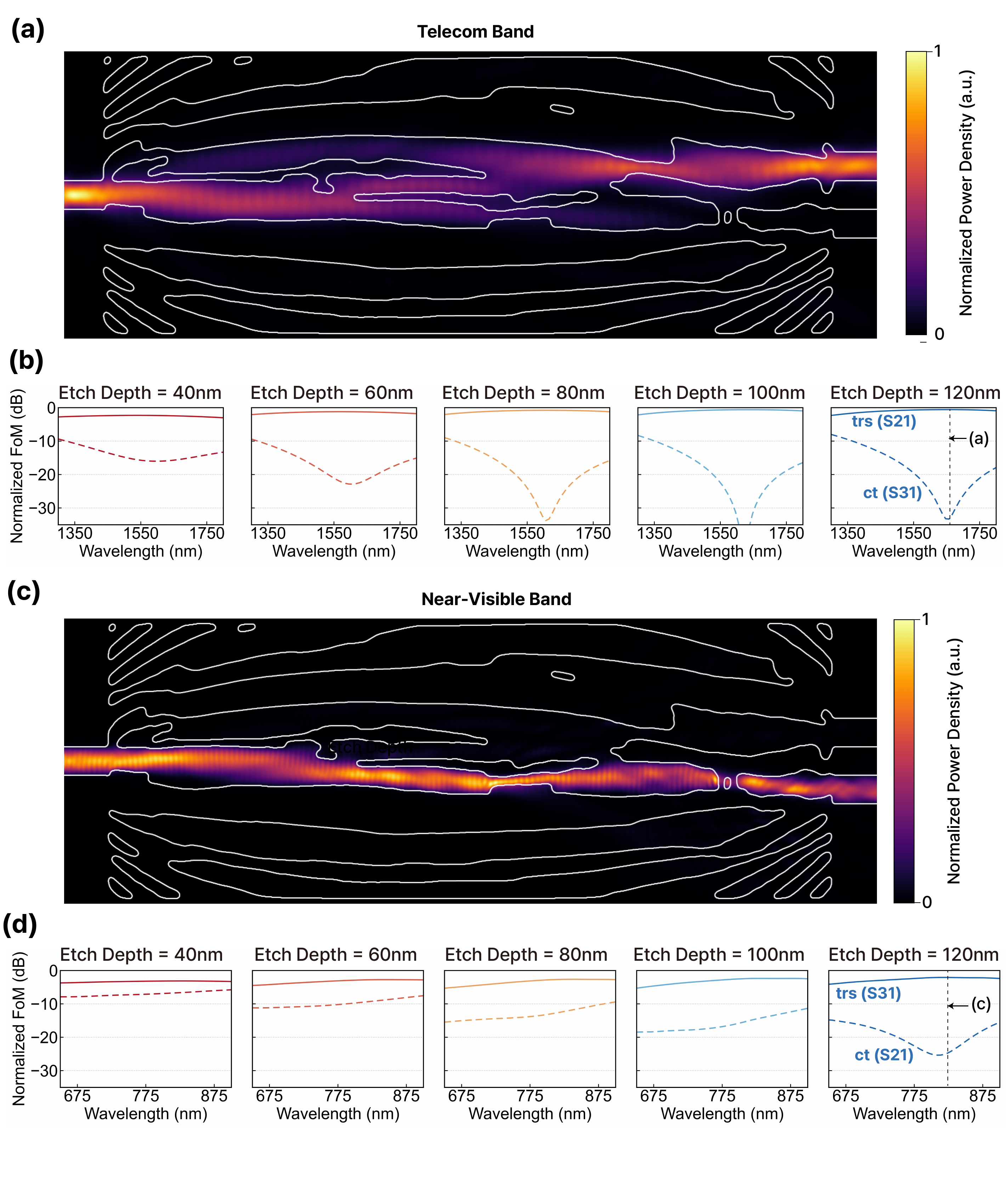}}
\caption{\fontsize{9pt}{7.5pt}\selectfont
Verification of the device performance using three-dimensional finite-difference time-domain simulations. (a) and (b) present the results for the telecom band, whereas (c) and (d) show the results for the near-visible band. The quantity visualized in (a) and (c) is the normalized electromagnetic energy density, $U=\varepsilon_0\varepsilon_r|\mathbf{E}|^2+\mu_0|\mathbf{H}|^2$, normalized to the range from 0 to 1. The field maps are shown at 1650 and 825~nm, corresponding to the transmission maxima of the three-dimensional structure, which are red-shifted from the design bands by the residual approximation of the BEI model. The maps correspond to an etch depth of 120~nm, which exhibits the highest signal-to-crosstalk ratio among the evaluated etch depths, as determined from the $S_{21}$ and $S_{31}$ spectra presented in (b) and (d).\label{fig4}}
\end{figure*}

Figure~\ref{fig4} presents the three-dimensional FDTD verification of the optimized demultiplexer for etch depths from 40 to 120~nm. The power-density distributions in Figs.~\ref{fig4}(a) and \ref{fig4}(c) correspond to the 120-nm etch depth at the wavelengths marked by the dashed lines in Figs.~\ref{fig4}(b) and \ref{fig4}(d), respectively. In the telecom band, incidence light is routed predominantly to the upper output waveguide, whereas the near-visible input is directed to the lower output waveguide. The field distributions confirm wavelength-selective routing through the same inverse-designed structure with limited power coupled to the undesired ports.

Figures~\ref{fig4}(b) and \ref{fig4}(d) show the target-port transmission and crosstalk spectra for each etch depth. As the etch depth increases, the wavelength-selective response becomes more pronounced. The target transmission remains relatively high, while the crosstalk minimum becomes deeper. At an etch depth of 120~nm, the crosstalk is suppressed by more than 30~dB in the telecom band and by approximately 25~dB in the near-visible band near the selected operating wavelengths. These results indicate that the stronger lateral index contrast produced by deeper etching improves separation between the desired and undesired output channels.

The transmission and crosstalk spectra shift with etch depth, indicating that the vertical geometry affects the spectral response of the device. Although the three-dimensional structure shows a spectral shift relative to the BEI prediction, the 120-nm-deep design maintains efficient routing and strong crosstalk suppression in both wavelength bands. These full-wave simulations therefore verify that the BEI-optimized structure remains effective after reconstruction into the three-dimensional TFLN geometry.

\begin{figure*}[h]
\centerline{\includegraphics[width=\linewidth]{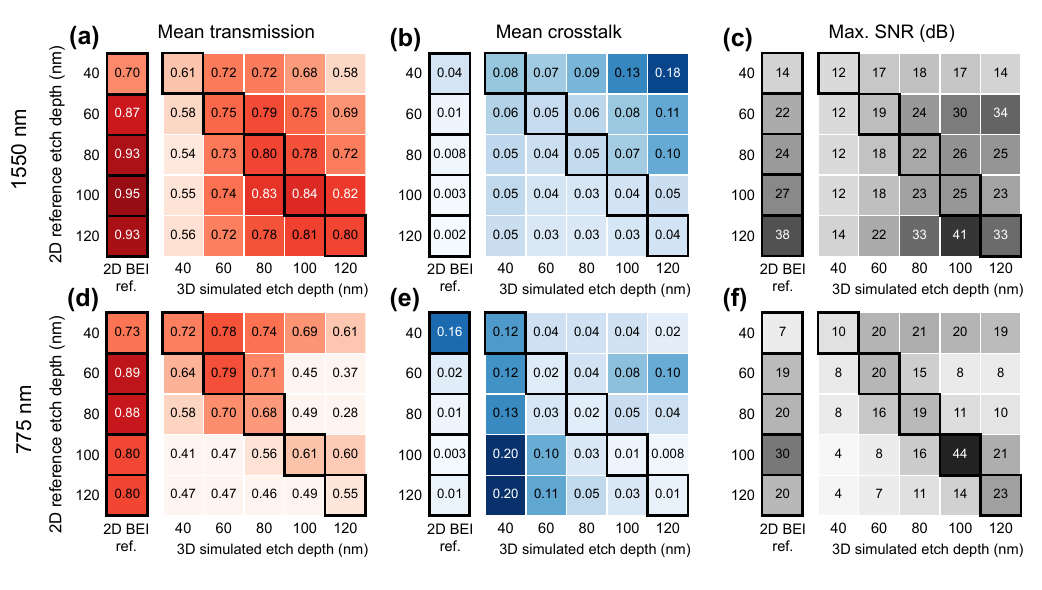}}
\caption{\fontsize{9pt}{7.5pt}\selectfont
Comparison between the optimization results obtained using the two-dimensional BEI model and the corresponding three-dimensional verification results. (a)–(c) summarize the results for the 1550 nm band, whereas (d)–(f) present those for the 775 nm band. (a) and (d) show the transmission to the designated output port, (b) and (e) show the crosstalk to the undesired output port, and (c) and (f) show the maximum signal-to-noise ratio. The three-dimensional result obtained at the reference etch depth targeted in the corresponding two-dimensional BEI optimization is highlighted by a black square.\label{fig5}}
\end{figure*}

Figure~\ref{fig5} compares the BEI predictions with full three-dimensional FDTD results across different reference and reconstructed etch depths. In each panel, the leftmost column shows the two-dimensional BEI result for a design optimized at the specified reference etch depth, while the remaining columns show the corresponding three-dimensional responses for reconstructed etch depths from 40 to 120~nm. The black squares indicate the cases in which the reconstructed three-dimensional etch depth matches the reference value used during BEI optimization.

For the 1550-nm band, the three-dimensional simulations reproduce the overall performance trend predicted by the BEI model, as shown in Figs.~\ref{fig5}(a)--\ref{fig5}(c). Along the matched-depth diagonal, the mean target transmission increases from 0.61 at an etch depth of 40~nm to 0.84 at 100~nm, while the mean crosstalk decreases from 0.08 to 0.04. The maximum signal-to-noise ratio increases from 12~dB at 40~nm to 25~dB at 100~nm and reaches 33~dB for the 120-nm design. These results show that the BEI-optimized structures retain high target transmission and strong crosstalk suppression after reconstruction into the full three-dimensional geometry.

Figures~\ref{fig5}(d)--\ref{fig5}(f) show the corresponding results for the 775-nm band. Along the matched-depth diagonal, the mean target transmission is 0.72, 0.79, 0.68, 0.61, and 0.55 for etch depths from 40 to 120~nm, respectively. Although the target transmission decreases for deeper etches, the mean crosstalk is reduced to 0.02 at 60 and 80~nm and to 0.01 at 100 and 120~nm. The 100-nm design provides the strongest channel separation, with a maximum signal-to-noise ratio of 44~dB in the corresponding three-dimensional simulation, while the 120-nm design maintains a maximum signal-to-noise ratio of 23~dB.

The off-diagonal entries indicate the sensitivity of each optimized design to variations in etch depth. Several structures maintain low crosstalk over neighboring etch depths, although the target transmission and signal-to-noise ratio change with the vertical geometry. The two-dimensional BEI model generally predicts higher transmission and lower crosstalk than the corresponding three-dimensional simulations, particularly in the 775-nm band. Despite these differences, the three-dimensional results retain wavelength-selective routing and strong channel separation over suitable etch depths. These results support the BEI method as an efficient reduced-dimensional approach for designing birefringent TFLN devices, with three-dimensional simulations used to evaluate the remaining discrepancy associated with the full etched geometry.

\section{Experimental Results}

\subsection{Experimental Demonstration of Dual-Band Wavelength Demultiplexing}

To experimentally validate the BEI-based inverse-design approach, the optimized demultiplexer was fabricated on X-cut TFLN and characterized in both the telecom and near-visible wavelength bands. The measurements were used to evaluate whether the wavelength-selective routing predicted by the reduced-dimensional design and three-dimensional verification is preserved in the fabricated device.

\begin{figure*}[tp]
\centerline{\includegraphics[width=\linewidth]{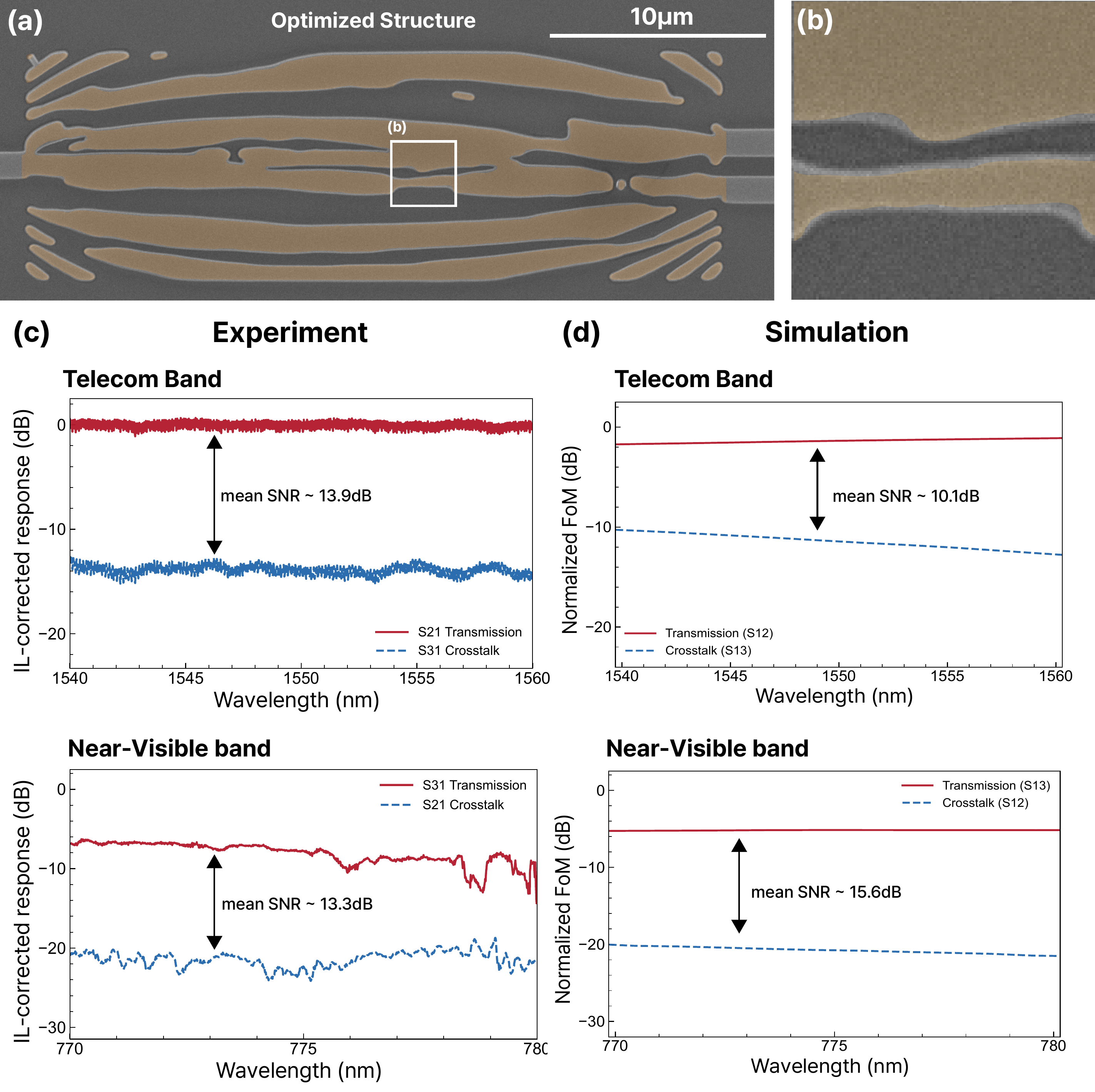}}
\caption{\fontsize{9pt}{7.5pt}\selectfont
Fabrication and experimental characterization of the optimized LN WDM. (a) Scanning electron microscopy (SEM) image of the fabricated device. (b) Magnified SEM image, confirming that the fabricated geometry closely matches the intended design. (c) Measured spectral responses in the 1540–1560 nm telecom band (top) and the 770–780 nm near-visible band (bottom). The spectra were calibrated by excluding the facet-to-facet insertion loss of the reference waveguide. The transmission to the designated output port is shown by the red solid line, whereas the crosstalk to the undesired output port is shown by the blue dashed line. (d) Simulated spectral responses obtained using the geometry reconstructed from the fabricated device.\label{fig6}}
\end{figure*}

Figure~\ref{fig6} presents the fabricated device and its measured dual-band demultiplexing response. The top-view SEM image in Fig.~\ref{fig6}(a) shows that the inverse-designed pattern was transferred across the full $30\times10~\mu\mathrm{m}^2$ design region with good fidelity. The enlarged image in Fig.~\ref{fig6}(b) further shows the narrow LN features and small voids required by the optimized geometry.

The fabricated device in Fig.~\ref{fig6} and the corresponding simulations use an etch depth of 180~nm, as confirmed by the cross-sectional profile in Fig.~S15, rather than the 120~nm depth considered in the preceding numerical analysis. Although simulations indicated that a 120~nm etch was sufficient for wavelength-selective routing within the inverse-designed region, the experimental device required additional waveguide bends to spatially separate the output ports. Simulations of this configuration showed substantial slab-mode leakage at the shallower etch depth, as shown in Fig.~S19. A deeper etch of 180~nm was therefore used to improve optical confinement and reduce bending loss.

The device was characterized in the telecom and near-visible bands using TSL-550 and CTL~780 tunable lasers, respectively. Light was launched into input port~1, and the spectra at output ports~2 and 3 were recorded. In the telecom band, $S_{21}$ denotes the desired transmission and $S_{31}$ the crosstalk, while these roles are reversed in the near-visible band. The measurement setup is shown in Fig.~S18. The measured spectra were normalized to a reference waveguide to compensate for coupling and common-path losses.

The measured spectra in Fig.~\ref{fig6}(c) were corrected by subtracting reference coupling losses of 10.27~dB and 15.06~dB for the telecom and near-visible bands, respectively. In the telecom band, the desired $S_{21}$ response exceeds the $S_{31}$ crosstalk by an average of 13.9~dB over 1540--1560~nm. In the near-visible band, the desired $S_{31}$ response exceeds the $S_{21}$ crosstalk by an average of 13.3~dB over 770--780~nm. These measurements demonstrate wavelength-selective routing between the telecom and near-visible bands in a compact TFLN demultiplexer.

To compare the measurements with full-wave simulations, the fabricated geometry was reconstructed from the SEM images and evaluated numerically. As shown in Fig.~\ref{fig6}(d), the simulated mean signal-to-crosstalk ratios are 10.1~dB in the telecom band and 15.6~dB in the near-visible band. Compared with these predictions, the measured device shows stronger channel separation in the telecom band and slightly weaker separation in the near-visible band.

The remaining discrepancy may arise from fabrication-induced variations in linewidth, sidewall geometry, and surface roughness. For a given dimensional deviation, the relative perturbation is larger at shorter wavelengths, making the near-visible response more sensitive to fabrication errors. Sidewall roughness can also produce stronger scattering losses at shorter wavelengths. Because these effects were not independently quantified, the measurement-simulation discrepancy cannot be attributed to a single mechanism. Nevertheless, the fabricated device maintains the intended port selectivity in both wavelength bands, supporting the experimental feasibility of the BEI-based inverse-design method for TFLN wavelength demultiplexers.

\subsection{Experimental Characterization of Cascaded Wavelength Demultiplexers}

\begin{figure*}[tp]
\centerline{\includegraphics[width=\linewidth]{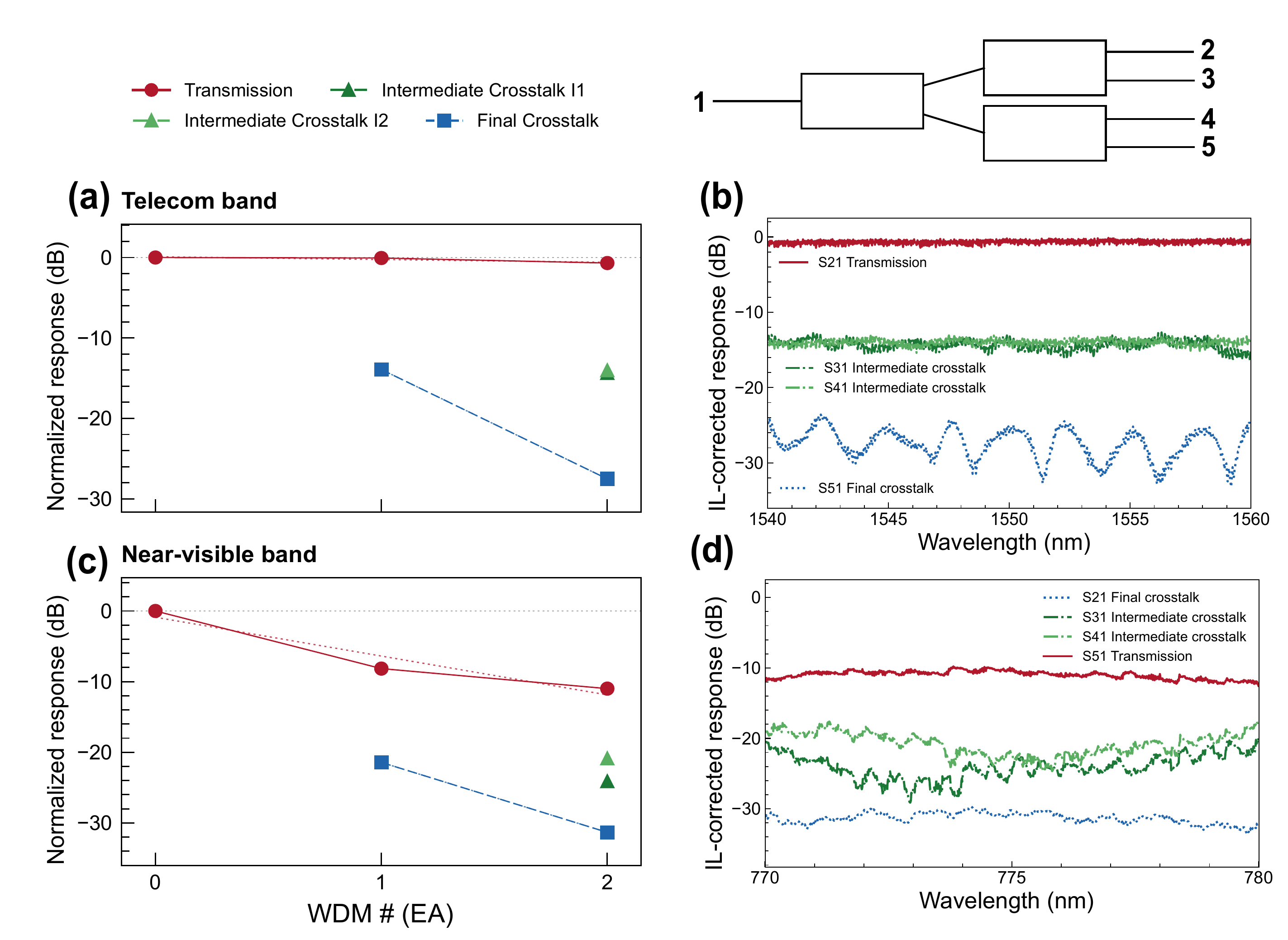}}
\caption{\fontsize{9pt}{7.5pt}\selectfont
Measured performance of the cascaded WDM devices. To isolate the intrinsic insertion loss of the WDM devices, reference coupling losses of $10.27$ dB (telecom band) and $15.06$ dB (near-visible band) were subtracted from all measured data in (a)-(d). (a) Cumulative insertion loss in the telecom band as a function of the number of cascaded WDM devices. The linear fit yields a slope of $-0.33$ dB/device. (b) Transmission ($S_{21}$), intermediate crosstalks($S_{31}$, $S_{41}$) and crosstalk ($S_{51}$) spectra in the telecom band (1550 nm). An extinction ratio of  26.8 dB is achieved, calculated based on the band-averaged value. (c) Cumulative insertion loss in the near-visible band as a function of the number of cascaded WDM devices, with a fitted slope of $-5.13$ dB/device. (d) Transmission ($S_{51}$), intermediate crosstalks($S_{41}$, $S_{31}$) and crosstalk ($S_{21}$) spectra in the near-visible band (775 nm), exhibiting an extinction ratio of 20.3 dB based on the band-averaged values.
\label{fig7}}
\end{figure*}

The cascaded configuration was evaluated to determine whether serially connected WDM devices provide stronger channel isolation than a single device. Figure~\ref{fig7}(a) shows the telecom-band response as the number of cascaded stages increases. The desired transmission decreases gradually with each additional stage, while the final crosstalk is suppressed more strongly. As shown in Fig.~\ref{fig7}(b), $S_{21}$ remains the dominant output over 1540--1560~nm, whereas the final $S_{51}$ crosstalk is strongly attenuated. The cascaded device achieves a band-averaged extinction ratio of approximately 26.8~dB, demonstrating improved channel isolation in the telecom band.

Figures~\ref{fig7}(c) and \ref{fig7}(d) show the corresponding results for the near-visible band. The desired $S_{51}$ response remains higher than both the intermediate and final crosstalk signals over 770--780~nm. After cascading, the final $S_{21}$ crosstalk is suppressed by approximately 20.3~dB relative to the desired transmission based on the band-averaged measurements. These results show that the cascaded configuration maintains the intended wavelength routing while improving channel isolation in both spectral bands.

The intermediate crosstalk spectra provide insight into the leakage paths through the cascaded device. In the telecom band, the $S_{31}$ and $S_{41}$ responses are similar, indicating comparable attenuation along the two intermediate leakage paths. In the near-visible band, however, $S_{31}$ is lower than $S_{41}$, suggesting greater attenuation along one of the leakage paths. This difference may arise from the higher sensitivity of shorter-wavelength light to modal mismatch, dimensional variations, and scattering, which can increase propagation and radiation losses.

The transmission trends in Figs.~\ref{fig7}(a) and \ref{fig7}(c) were used to estimate the incremental insertion loss of the cascaded devices. Linear fitting yields 0.33~dB per device in the telecom band and 5.13~dB per device in the near-visible band. The lower telecom-band loss indicates more efficient propagation through successive WDM stages, whereas the higher near-visible loss is consistent with the greater sensitivity of shorter-wavelength light to sidewall roughness, geometric variations, and mode conversion. Despite this difference, the cascaded devices preserve clear port selectivity and provide improved crosstalk suppression in both spectral bands. Measurements for additional geometric constraints and target etch depths are summarized in Fig.~S20, with representative spectra shown in Figs.~S21--S23.

\section{Conclusion}

We demonstrated a reduced-dimensional inverse-design method for birefringent TFLN photonic devices that avoids the computational burden of three-dimensional optimization. The BEI method represents the in-plane anisotropic optical response of X-cut LN using effective-index tensors derived from slab modes, while the mid-plane geometry and fabrication constraints account for the slanted sidewalls produced by LN etching. This treatment converts the anisotropic three-dimensional design problem into a two-dimensional optimization problem while preserving the optical characteristics required for wavelength routing. For the demonstrated device, the reduced model decreased the forward-simulation time by approximately 711-fold under identical computational resources.

Using this approach, we designed and fabricated a compact $30\times10~\mu\mathrm{m}^2$ demultiplexer that separates the 1550- and 775-nm bands. Three-dimensional simulations across multiple etch depths verified that the BEI-optimized structures retain wavelength-selective routing after reconstruction into the full etched geometry. Experimentally, the fabricated device achieved mean signal-to-crosstalk ratios of 13.9~dB over 1540--1560~nm and 13.3~dB over 770--780~nm. Cascading two devices further increased the band-averaged extinction ratios to 26.8~dB and 20.3~dB, respectively.

These results establish that polarization-dependent effective-index modeling can substantially reduce the computational cost of inverse design without requiring three-dimensional optimization at every iteration. Although full three-dimensional verification remains important for evaluating vertical radiation, sidewall geometry, and spectral shifts, the BEI method enables computationally efficient exploration of large design spaces before such detailed simulations are performed. This capability can facilitate the design of compact multiwavelength components for nonlinear and quantum photonic circuits and may be extended to other birefringent materials through appropriate effective-index tensors.

\section{Methods}

\subsection{Birefringent Effective Index Method}

Slab and waveguide eigenmodes used to construct and validate the BEI model were calculated using MIT Photonic Bands~\cite{johnson_block-iterative_2001}. The relative permittivity tensor of X-cut LN was defined with the extraordinary axis along the $y$-direction. A spatial resolution of 100~pixels/$\mu$m was used for the waveguide eigenmode calculations to resolve the slanted sidewalls.

For a 300-nm-thick LN film, slab-mode effective indices were calculated for etch depths from 40 to 120~nm using a 5-$\mu$m-wide computational domain along the $z$-direction. The calculations were performed at the center wavelengths of 1550 and 775~nm for in-plane propagation along the $x$- and $y$-directions. The LN permittivity tensors were defined as $\varepsilon_{\mathrm{LN},1550}=(\varepsilon_x,\varepsilon_y,\varepsilon_z)=(4.873,4.585,4.873)$ and $\varepsilon_{\mathrm{LN},775}=(5.109,4.752,5.109)$.

The SiO$_2$ substrate and cladding were modeled using $\varepsilon_{\mathrm{SiO_2}}=2.085$, with their weak dispersion neglected. The resulting propagation-direction- and wavelength-dependent effective indices were used to construct the equivalent permittivity tensors of the two-dimensional BEI model. Dispersion between the two operating bands was therefore captured through separate effective-index tensors, while the effective indices within each optimization band were treated as nondispersive because the refractive-index variation over each bandwidth is small.

To evaluate the BEI approximation, reference waveguide eigenmodes were calculated using the full cross-sectional geometries. The waveguide cross sections were modeled in $5~\mu\mathrm{m}\times5~\mu\mathrm{m}$ computational domains in the $xz$- and $yz$-planes for propagation along the $y$- and $x$-directions, respectively. The resulting modal effective indices were compared with those obtained from the reduced-dimensional BEI model.

\subsection{Simulation Methods}

Two-dimensional FDTD simulations and inverse design were performed using Meep~\cite{oskooi_meep_2010} on a workstation equipped with an Intel Core i7-12700K processor and 64~GB of RAM, using 10 of the 12 physical CPU cores. Three-dimensional validation simulations were performed on a separate workstation equipped with two 64-core AMD EPYC 9554 processors and 1.5~TB of RAM, using 120 of the 128 physical CPU cores. To provide a direct computational-cost comparison, the two-dimensional and three-dimensional forward simulations were also executed on the same Intel-based workstation under identical hardware and parallelization settings. The compared simulations used the corresponding two-dimensional and three-dimensional representations of the same device, so that the runtime difference primarily reflects the dimensional reduction introduced by the BEI approximation.

\subsubsection{Simulation Settings and Design Region}

For the two-dimensional inverse-design simulations, the input waveguide was excited by Gaussian pulses centered at 1550 and 775~nm, each with a fractional bandwidth of $\Delta f/f_0=0.1$. The input and output access waveguides extended $2~\mu\mathrm{m}$ from the design region. A $1~\mu\mathrm{m}$ spacing was included between the device and the upper and lower boundaries to reduce boundary interactions, and the computational domain was terminated by $1~\mu\mathrm{m}$-thick perfectly matched layers.

The material distribution within the design region was represented by a continuous design field $\rho(\mathbf{x})\in[0,1]$, where $\mathbf{x}=(x,y)$ denotes the in-plane position. The relative permittivity tensor at each position was interpolated between the residual etched slab and the unetched ridge as

\begin{equation}
\boldsymbol{\epsilon}_{r}^{(k)}(\rho(\mathbf{x}))
=
\left(1-\rho(\mathbf{x})\right)
\boldsymbol{\epsilon}_{\mathrm{sub}}^{(k)}
+
\rho(\mathbf{x})
\boldsymbol{\epsilon}_{\mathrm{str}}^{(k)},
\label{eq:material_interp}
\end{equation}

where the material tensors are defined as

\begin{equation}
\boldsymbol{\epsilon}_{m}^{(k)}
=
\mathrm{diag}\!\left[
\left(n^{\mathrm{eff}}_{m,o,k}\right)^2,
\left(n^{\mathrm{eff}}_{m,e,k}\right)^2,
\left(n^{\mathrm{eff}}_{m,o,k}\right)^2
\right],
\qquad
m\in\{\mathrm{sub},\mathrm{str}\}.
\label{eq:material_tensors}
\end{equation}

Here, $k\in\{1550,775\}$ denotes the telecom or near-visible wavelength band. The subscripts $\mathrm{sub}$ and $\mathrm{str}$ indicate the residual etched slab and unetched ridge regions, respectively, while $o$ and $e$ denote the ordinary and extraordinary effective indices. The tensor components correspond to the simulation coordinates $(x,y,z)$, with the extraordinary axis aligned along $y$. The conditions $\rho(\mathbf{x})=0$ and $\rho(\mathbf{x})=1$ represent the etched and unetched regions, respectively. Intermediate values provide a continuous material interpolation during optimization and are progressively driven toward a binary distribution using the filtering and projection procedures described below.

\subsubsection{Figure of Merit}

The time-averaged power through a monitoring plane $A$ was calculated as

\begin{equation}
P_{\alpha}
=
\frac{1}{2}\mathrm{Re}
\left[
\int_A
\left(
\mathbf{E}_{\alpha}
\times
\mathbf{H}_{\alpha}^{*}
\right)
\cdot\hat{\mathbf{n}}\,dA
\right],
\quad
\alpha\in\{\mathrm{out},\mathrm{src},\mathrm{tar}\},
\label{eq:Palpha}
\end{equation}
where $\hat{\mathbf{n}}$ is the unit vector normal to the monitoring plane. The incident power $P_{\mathrm{src}}$ and target-mode fields were obtained from separate normalization simulations of an isolated straight waveguide, with the monitoring plane located $0.5~\mu\mathrm{m}$ from the source. The output monitoring planes were positioned at the centers of the output access waveguides, $1~\mu\mathrm{m}$ from the design-region boundary. Here, $P_{\mathrm{src}}$, $P_{\mathrm{out}}$, and $P_{\mathrm{tar}}$ denote the powers associated with the incident field, output field, and target waveguide mode, respectively.

The total transmission through the selected output plane was defined as

\begin{equation}
\tau
=
\frac{P_{\mathrm{out}}}{P_{\mathrm{src}}},
\label{eq:tau}
\end{equation}
which includes all transmitted field components regardless of their modal composition. The normalized overlap with the target waveguide mode was evaluated as

\begin{equation}
\xi
=
\frac{
\left|
\int_A
\left[
\mathbf{E}_{\mathrm{out}}
\times
\mathbf{H}_{\mathrm{tar}}^{*}
\right]
\cdot\hat{\mathbf{n}}\,dA
\right|^2
}{
4P_{\mathrm{out}}P_{\mathrm{tar}}
},
\label{eq:overlap}
\end{equation}
where $\xi$ quantifies the coupling of the transmitted field to the target mode~\cite{snyder_decomposition_1983}. The figure of merit was then defined as

\begin{equation}
\mathcal{T}
=
\tau\xi,
\label{eq:fom}
\end{equation}
which represents the fraction of incident power coupled to the target output mode. Under the adopted normalization and for a passive device, $\tau$, $\xi$, and $\mathcal{T}$ range from 0 to 1. The value of $\mathcal{T}$ was evaluated at both output ports and used to optimize wavelength-selective routing.

\subsubsection{Maxwell-Constrained Minimax Optimization}

Using the band-dependent permittivity tensor defined in Eq.~\eqref{eq:material_interp}, the electric field at each sampled wavelength satisfies the frequency-domain Maxwell equation

\begin{equation}
\nabla\times
\frac{1}{\mu_0\mu_r}
\nabla\times\mathbf{E}_{k,p}
-
\omega_{k,p}^{2}\varepsilon_0
\boldsymbol{\epsilon}_{r}^{(k)}(\rho)
\mathbf{E}_{k,p}
=
-i\omega_{k,p}\mathbf{J}_{k,p}.
\label{eq:maxwell_constraint}
\end{equation}

Here, $\rho$ denotes the design field over the design region. The index $k\in\{1550,775\}$ identifies the telecom or near-visible band, and $p=1,\ldots,5$ denotes one of the five wavelength samples $\lambda_{k,p}$ within each band. The corresponding angular frequency is $\omega_{k,p}=2\pi c_0/\lambda_{k,p}$, where $c_0$ is the speed of light in vacuum. The quantities $\mathbf{E}_{k,p}$ and $\mathbf{J}_{k,p}$ denote the electric field and impressed source current at $\omega_{k,p}$, respectively. The materials were assumed to be nonmagnetic with $\mu_r=1$.

The target-transmission and crosstalk errors at each sampled wavelength were defined as

\begin{equation}
e_{k,p}^{\mathrm{tar}}(\rho)
=
1-\mathcal{T}_{k,p}^{\mathrm{tar}}(\rho),
\qquad
e_{k,p}^{\mathrm{ct}}(\rho)
=
\mathcal{T}_{k,p}^{\mathrm{ct}}(\rho),
\label{eq:routing_errors}
\end{equation}
where $\mathcal{T}_{k,p}^{\mathrm{tar}}$ and $\mathcal{T}_{k,p}^{\mathrm{ct}}$ denote transmission into the desired and undesired output modes, respectively. The upper output port is designated as the target for the 1550-nm band, whereas the lower output port is the target for the 775-nm band. Accordingly, $e_{k,p}^{\mathrm{tar}}$ quantifies the target-transmission deficiency, while $e_{k,p}^{\mathrm{ct}}$ quantifies the crosstalk.

The two spectral bands, five wavelength samples per band, and target-transmission and crosstalk terms yield 20 competing error terms. Rather than combining them using fixed weights, the optimization was formulated in minimax form to reduce the largest error among all sampled conditions~\cite{hammond_photonic_2021}

\begin{equation}
\begin{aligned}
\min_{\rho,t}\quad
& t
\\
\textrm{s.t.}\quad
&
e_{k,p}^{\mathrm{tar}}(\rho)\leq t,
\\
&
e_{k,p}^{\mathrm{ct}}(\rho)\leq t,
\\
&
g_{\mathrm{LW}}(\rho)\leq0,
\qquad
g_{\mathrm{LS}}(\rho)\leq0,
\\
&
0\leq\rho\leq1,
\\
&
k\in\{1550,775\},
\quad
p=1,\ldots,5.
\end{aligned}
\label{eq:opt_problem}
\end{equation}

The auxiliary variable $t$ provides an upper bound on the target-transmission deficiency and crosstalk across all sampled wavelengths. Minimizing $t$ therefore improves the worst-performing routing condition and promotes uniform performance across both spectral bands. The constraints $g_{\mathrm{LW}}$ and $g_{\mathrm{LS}}$ impose the prescribed minimum linewidth and minimum spacing, respectively, as defined below. The optimization problem was solved using the method of moving asymptotes~\cite{svanberg_class_2002} implemented in the Python NLopt package~\cite{johnson_nlopt_2007}.

\subsubsection{Conic Filter and Projection}

A conic density filter and a smooth projection were applied to suppress subresolution variations and promote a binary geometry~\cite{hammond_photonic_2021,wang_projection_2011}. The filtered design field was defined as

\begin{equation}
\widetilde{\rho}(\mathbf{x})
=
\int_{\Omega_{\mathrm{D}}}
w(\mathbf{x},\mathbf{x}')
\rho(\mathbf{x}')\,d\mathbf{x}',
\label{eq:conic_filter}
\end{equation}
where $\Omega_{\mathrm{D}}$ denotes the design region. The normalized conic kernel was given by

\begin{equation}
w(\mathbf{x},\mathbf{x}')
=
\begin{cases}
\dfrac{1}{C(\mathbf{x})}
\left(
1-\dfrac{\lVert\mathbf{x}-\mathbf{x}'\rVert}{R}
\right),
&
\lVert\mathbf{x}-\mathbf{x}'\rVert\leq R,
\\[8pt]
0,
&
\lVert\mathbf{x}-\mathbf{x}'\rVert>R,
\end{cases}
\label{eq:conic_kernel}
\end{equation}
with the normalization factor

\begin{equation}
C(\mathbf{x})
=
\int_{\Omega_{\mathrm{D}}}
\max
\left(
0,
1-\frac{\lVert\mathbf{x}-\mathbf{x}'\rVert}{R}
\right)
d\mathbf{x}'.
\label{eq:filter_normalization}
\end{equation}

The filter radius $R$ sets the spatial scale of the convolution: variations substantially smaller than $R$ are smoothed by the averaging, which discourages subresolution features. In this work, $R$ was set equal to the mid-plane minimum length scale of each geometric-constraint condition (Table~S1). To prevent perturbation of the optimized geometry when the minimum feature-size and minimum void-size constraints are activated, the same radius $R$ was used in these constraints, described below.

The filtered field was then mapped toward a binary material distribution using the smooth projection.

\begin{equation}
\bar{\rho}(\mathbf{x})
=
\frac{
\tanh(\beta\eta)
+
\tanh\!\left[
\beta\left(
\widetilde{\rho}(\mathbf{x})-\eta
\right)
\right]
}{
\tanh(\beta\eta)
+
\tanh\!\left[
\beta(1-\eta)
\right]
},
\label{eq:density_projection}
\end{equation}
where $\eta$ is the projection threshold and $\beta$ controls the projection sharpness. Increasing $\beta$ drives $\bar{\rho}$ toward 0 or 1 while retaining a differentiable mapping during optimization.

In this work, $\eta=0.5$ was used, consistent with the symmetric thresholds. To allow the design to converge gradually and avoid poor local optima, the projection sharpness followed a continuation scheme initialized at $\beta=2$ and doubled every 50 iterations up to $\beta=2^{11}$.

\subsubsection{Geometric Constraints}
For the minimum-linewidth (LW) constraint, the solid-region indicator was defined as
\begin{equation}
I_{i}^{\mathrm{LW}}(\rho)
=
\bar{\rho}_{i}
\exp\!\left(
-c\left\lVert\nabla\widetilde{\rho}_{i}\right\rVert^{2}
\right),
\label{eq:lw_indicator}
\end{equation}
where $\widetilde{\rho}_{i}$ and $\bar{\rho}_{i}$ denote the filtered and projected fields evaluated at pixel $i$, and $c$ is a hyperparameter governing the convergence of the constraint functions. The factor $\bar{\rho}_{i}$ selects pixels belonging to the solid region, while the exponential term reduces contributions near material interfaces, where $\lVert\nabla\widetilde{\rho}_{i}\rVert$ is large. Using this indicator, the minimum-linewidth constraint was written as
\begin{equation}
g_{\mathrm{LW}}(\rho)
=
\frac{1}{N_d}
\sum_{i\in\mathcal{I}}
I_{i}^{\mathrm{LW}}(\rho)
\left[
\min\left(
\widetilde{\rho}_{i}-\eta_{e},
0
\right)
\right]^{2},
\label{eq:lw_constraint}
\end{equation}
where $\mathcal{I}$ is the set of design pixels and $N_d$ is their number. The threshold $\eta_e$ defines the required filtered-density level within solid features. The penalty becomes nonzero when a pixel identified as solid has $\widetilde{\rho}_{i}<\eta_e$, indicating insufficient support from the surrounding solid region.
Similarly, the void-region indicator was defined as
\begin{equation}
I_{i}^{\mathrm{LS}}(\rho)
=
\left(
1-\bar{\rho}_{i}
\right)
\exp\!\left(
-c\left\lVert\nabla\widetilde{\rho}_{i}\right\rVert^{2}
\right),
\label{eq:ls_indicator}
\end{equation}
where LS denotes the minimum line spacing. The factor $1-\bar{\rho}_{i}$ selects pixels belonging to the void region, and the corresponding minimum-spacing constraint was expressed as
\begin{equation}
g_{\mathrm{LS}}(\rho)
=
\frac{1}{N_d}
\sum_{i\in\mathcal{I}}
I_{i}^{\mathrm{LS}}(\rho)
\left[
\min\left(
\eta_{d}-\widetilde{\rho}_{i},
0
\right)
\right]^{2}.
\label{eq:ls_constraint}
\end{equation}

Here, $\eta_d$ defines the required filtered-density level within void regions. This penalty becomes nonzero when a pixel identified as void has $\widetilde{\rho}_{i}>\eta_d$, indicating that the surrounding void region is insufficiently developed. The geometric constraints were imposed as $g_{\mathrm{LW}}\leq0$ and $g_{\mathrm{LS}}\leq0$, which require both nonnegative penalty functions to vanish~\cite{zhou_minimum_2015}. In this work, $\eta_e=0.75$, $\eta_d=0.25$, and $c=1\times10^{-7}$ were used, following Ref.~\cite{hammond_photonic_2021}. The constraints were activated after 500 iterations of density optimization and projection and retained for the final 100 iterations, resulting in a total of 600 iterations.

For a structure with slanted sidewalls, a pattern defined at the mid-plane acquires different linewidths and spacings at the top and bottom surfaces. The lateral displacement of each sidewall from the mid-plane to either surface is $(T/2)\cot\theta$, where $T$ is the etch depth and $\theta$ is the sidewall angle measured from the horizontal plane. The corresponding change in the total feature or void width is therefore $T\cot\theta$. To ensure a minimum dimension $X$ at the limiting top or bottom surface, the constraint imposed at the mid-plane was defined as

\begin{equation}
X_{\mathrm{mid}}
=
X
+
T\cot\theta
=
X
+
T\tan\left(\frac{\pi}{2}-\theta\right).
\label{eq:sidewall_correction}
\end{equation}

The correction term accounts for the total width variation produced by the two slanted sidewalls over the etch depth. Separate corrections were applied to the minimum linewidth and minimum spacing because the most restrictive plane differs between solid and void regions. Three nominal geometric-constraint values of 50, 75, and 100~nm were investigated. The 50-nm condition was applied without sidewall correction, whereas the 75- and 100-nm conditions incorporated the correction in Eq.~\eqref{eq:sidewall_correction}. The constraint definitions and corresponding optimization conditions are summarized in Fig.~S7 and Table~S1.

\subsection{Fabrication Methods}

Devices designed with geometric constraints of 50 and 100~nm were fabricated and characterized for different structural scales and cascade configurations. The fabrication process is illustrated in Fig.~S15. A negative-tone electron-beam resist, ma-N~2405, was spin-coated onto the lithium-niobate-on-insulator substrate, followed by a conductive E-spacer layer to suppress electron charging. The device patterns were defined by electron-beam lithography using a JBX-A9 system (JEOL). After exposure, the E-spacer was removed, and the resist was developed in CT-D1 to form the waveguide mask.

The resist pattern was transferred into the LN layer by inductively coupled plasma reactive-ion etching using a PlasmaPro~100 Cobra system (Oxford Instruments) with an Ar-based process. Redeposited material and etching residues were subsequently removed by treatment in KOH at $65\,^{\circ}\mathrm{C}$ for 40~min, followed by oxygen plasma ashing at 350~W for 5~min. A 2-$\mu\mathrm{m}$-thick SiO$_2$ upper cladding was then deposited by plasma-enhanced chemical vapor deposition at $370\,^{\circ}\mathrm{C}$. The fabricated devices were finally annealed at $520\,^{\circ}\mathrm{C}$ for 2~h in an oxygen atmosphere to improve the deposited-film quality.

\bibliography{Thesis_LNDEMUX_final}

\newpage


\end{document}



\begin{center}
{\Large \textbf{Supporting Information}}\\[2em]

{\LARGE \textbf{
Inverse-Designed Lithium Niobate Wavelength Demultiplexer via Birefringent Effective Index Method
}}\\[2em]

Chihyeon Kim$^{1}$,
Minho Choi$^{2,3}$,
Munseong Bae$^{1}$,
Hyounghan Kwon$^{2,4,*}$,
and Haejun Chung$^{1,3,*}$\\[1em]

{\small
$^{1}$Department of Electronic Engineering, Hanyang University, Seoul,South Korea\\
$^{2}$Center of Quantum Technology, Korea Institute of Science and Technology(KIST), Seoul, South Korea\\
$^{3}$Department of Artificial Intelligence Semiconductor Engineering, Hanyang University, Seoul, South Korea\\

$^{4}$Division of Quantum Information, KIST School, Korea University of Science and Technology, Seoul, South Korea\\
Corresponing authors :
$^{*}$\href{mailto:hyounghankwon@kist.re.kr}{hyounghankwon@kist.re.kr}
$^{*}$\href{mailto:haejun@hanyang.ac.kr}{haejun@hanyang.ac.kr}
}
\end{center}

\vspace{2em}


{\large\textbf{Contents}}\\[1em]

\begin{tabular}{@{}p{2.5em}p{0.86\linewidth}@{}}
\hyperref[sec:S1]{\large S1} &
\hyperref[sec:S1]{\large Thickness-dependent Effective index}\\[0.6em]

\hyperref[sec:S2]{\large S2} &
\hyperref[sec:S2]{\large Birefringent Effective Index method}\\[0.6em]

\hyperref[sec:S3]{\large S3} &
\hyperref[sec:S3]{\large Design Region Size Optimization}\\[0.6em]

\hyperref[sec:S4]{\large S4} &
\hyperref[sec:S4]{\large Geometric Constraints for Slanted Sidewall}\\[0.6em]

\hyperref[sec:S5]{\large S5} &
\hyperref[sec:S5]{\large Optimization Results for Different Geometric Constraints}\\[0.6em]

\hyperref[sec:S6]{\large S6} &
\hyperref[sec:S6]{\large Optimization Result Summary}\\[0.6em]

\hyperref[sec:S7]{\large S7} &
\hyperref[sec:S7]{\large Spectral Shift According to Etch depths}\\[0.6em]

\hyperref[sec:S8]{\large S8} &
\hyperref[sec:S8]{\large SEM Images}\\[0.6em]

\hyperref[sec:S9]{\large S9} &
\hyperref[sec:S9]{\large Fabrication Methods}\\[0.6em]

\hyperref[sec:S10]{\large S10} &
\hyperref[sec:S10]{\large Measurement Setup}\\[0.6em]

\hyperref[sec:S11]{\large S11} &
\hyperref[sec:S11]{\large Measurement Results}\\[0.6em]

\end{tabular}

\newpage


\phantomsection
\label{sec:S1}
{\large\textbf{S1. Thickness-dependent Effective index}}

\begin{figure}[h!]
    \centering
    \includegraphics[width=\linewidth]{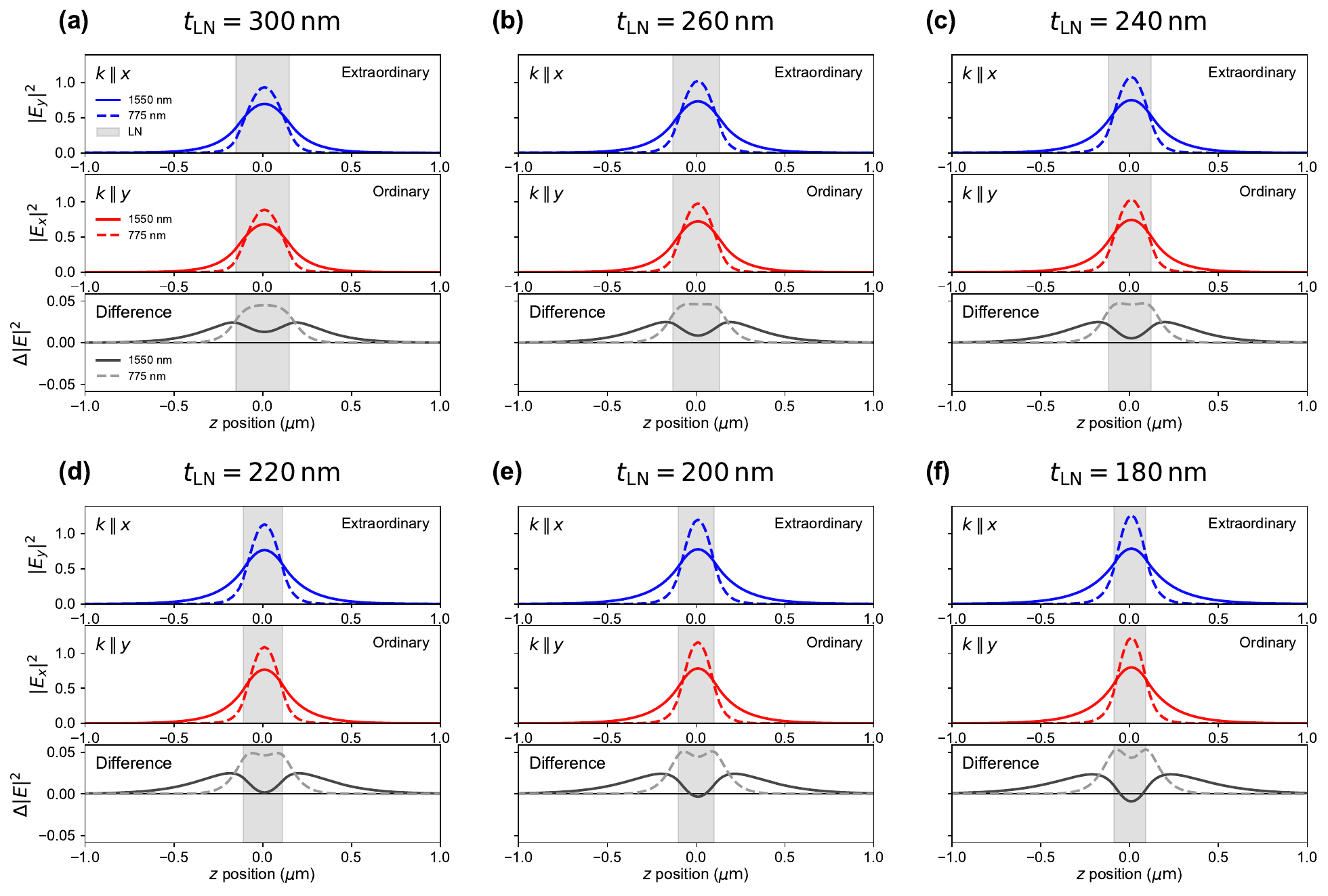}\
    \caption{%
        Electric-field distributions as a function of the lithium niobate thickness and propagation direction in x-cut LN. The LN thicknesses are (a) 300 nm, (b) 260 nm, (c) 240 nm, (d) 220 nm, (e) 200 nm, and (f) 180 nm. In each panel, the upper subpanel shows the intensity of $E_y$ field for propagation along the $x$ direction, the middle subpanel shows the intensity of $E_x$-field for propagation along the $y$ direction, and the lower subpanel shows the difference between the two field-intensity distributions. The results at 1550 nm and 775 nm are represented by solid and dashed lines, respectively.
    }
    \label{fig:S1}
\end{figure}

\clearpage


\phantomsection
\label{sec:S2}
{\large\textbf{S2. Birefringent Effective Index method}}

\begin{figure}[h!]
    \centering
    \includegraphics[width=\linewidth]{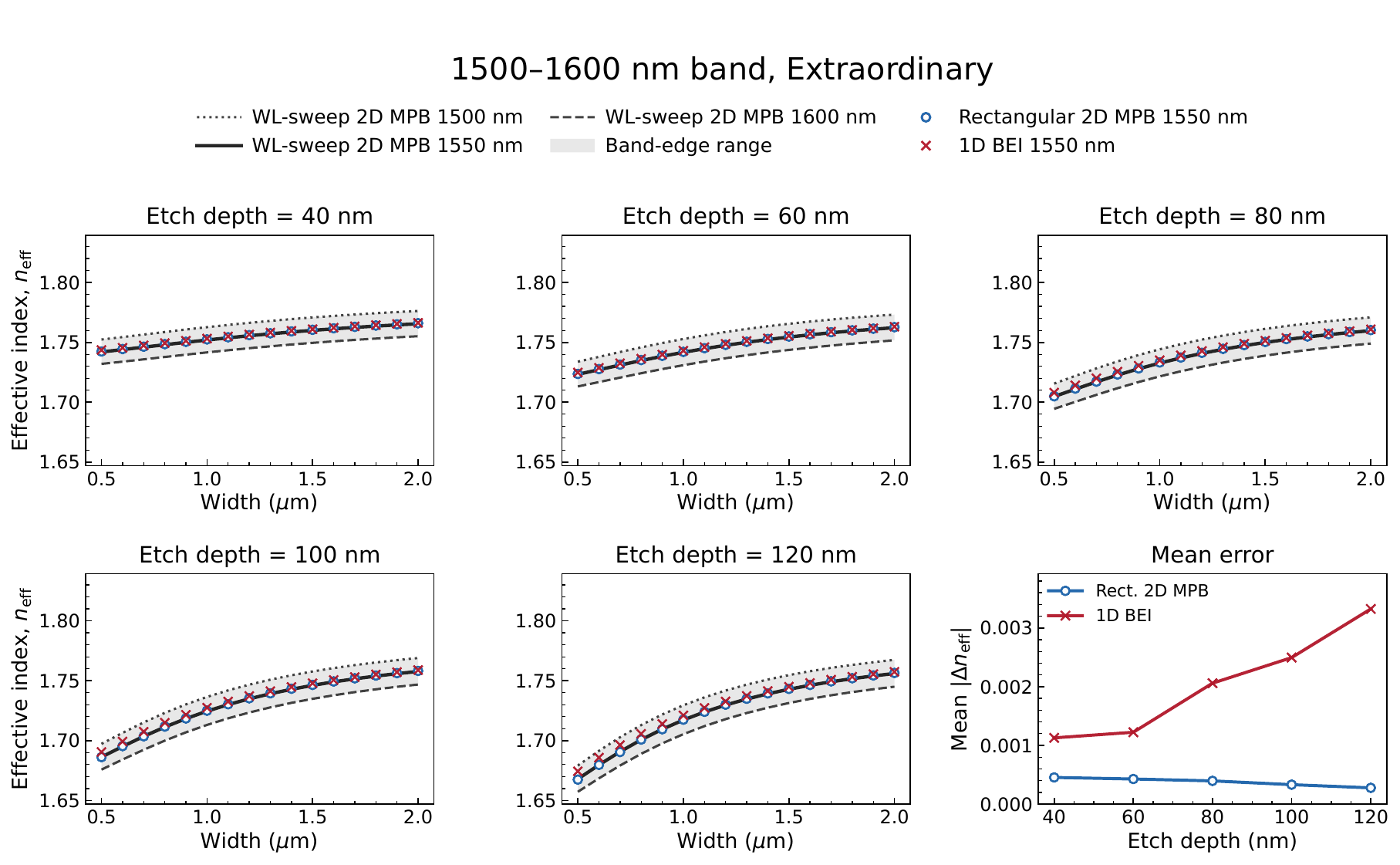}\
    \caption{%
        Comparison of the calculated effective indices at different etch depths for the extraordinary propagation direction in telecom band. From top to bottom, the panels show the results for etch depths of 40, 60, 80, 100, and 120 nm, respectively. The birefringent effective-index (BEI) values used in this work at 1550 nm are indicated by red crosses, whereas the results for rectangular waveguides with widths corresponding to the middle widths of the slanted sidewalls are indicated by blue circles. For the slanted-sidewall waveguides, the results at the center wavelength of the telecom band, 1550 nm, are shown by solid lines, while those at 1500 and 1600 nm are represented by finely and coarsely dashed lines, respectively. The corresponding variation across the telecom band is indicated by the gray shaded region. The lower-right panel shows the mean errors in the effective index obtained using the different calculation methods at the center wavelength.
    }
    \label{fig:S2}
\end{figure}

\begin{figure}[h!]
    \centering
    \includegraphics[width=\linewidth]{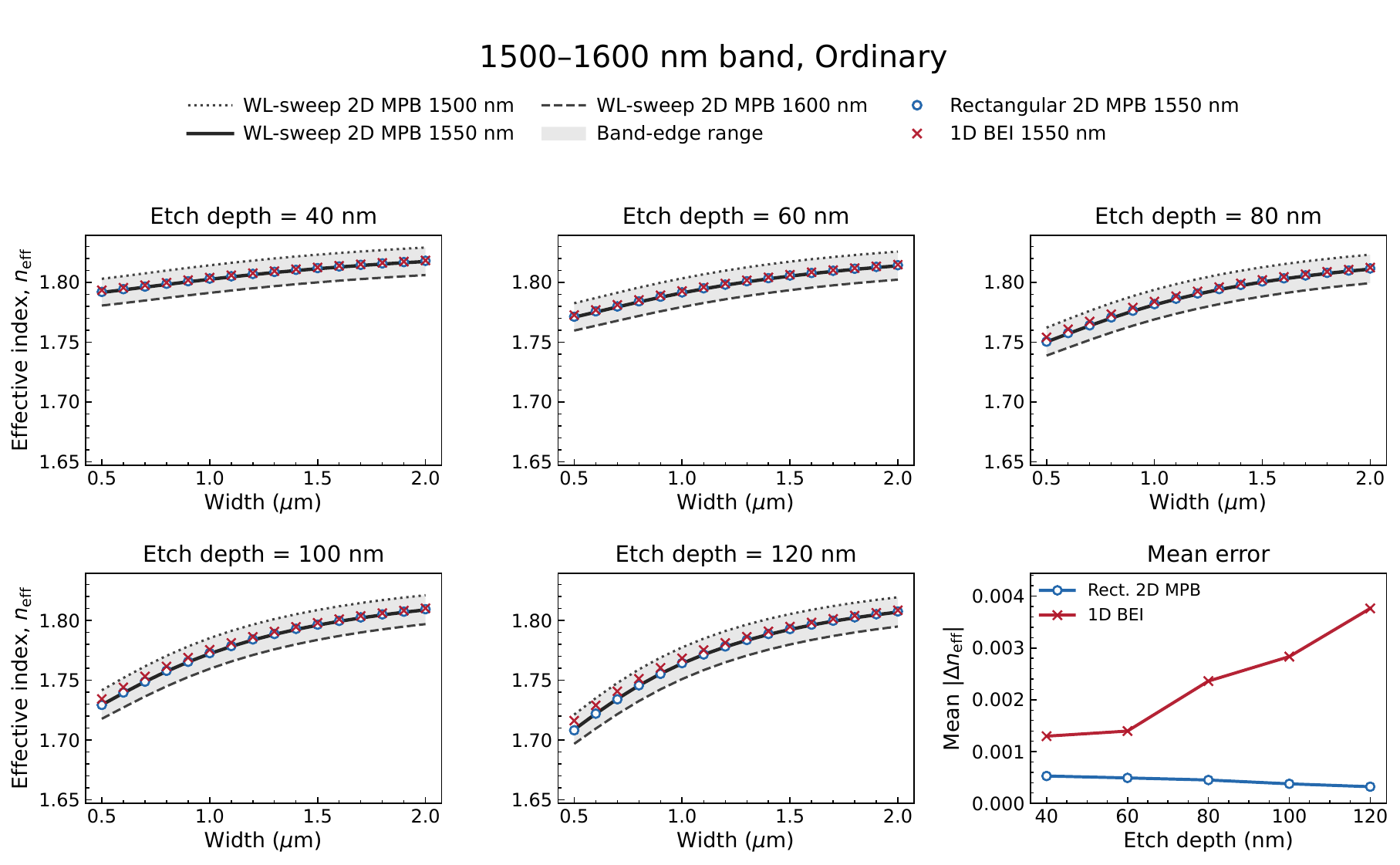}\
    \caption{%
        Comparison of the calculated effective indices at different etch depths for the ordinary propagation direction in telecom band.
    }
    \label{fig:S3}
\end{figure}

\begin{figure}[h!]
    \centering
    \includegraphics[width=\linewidth]{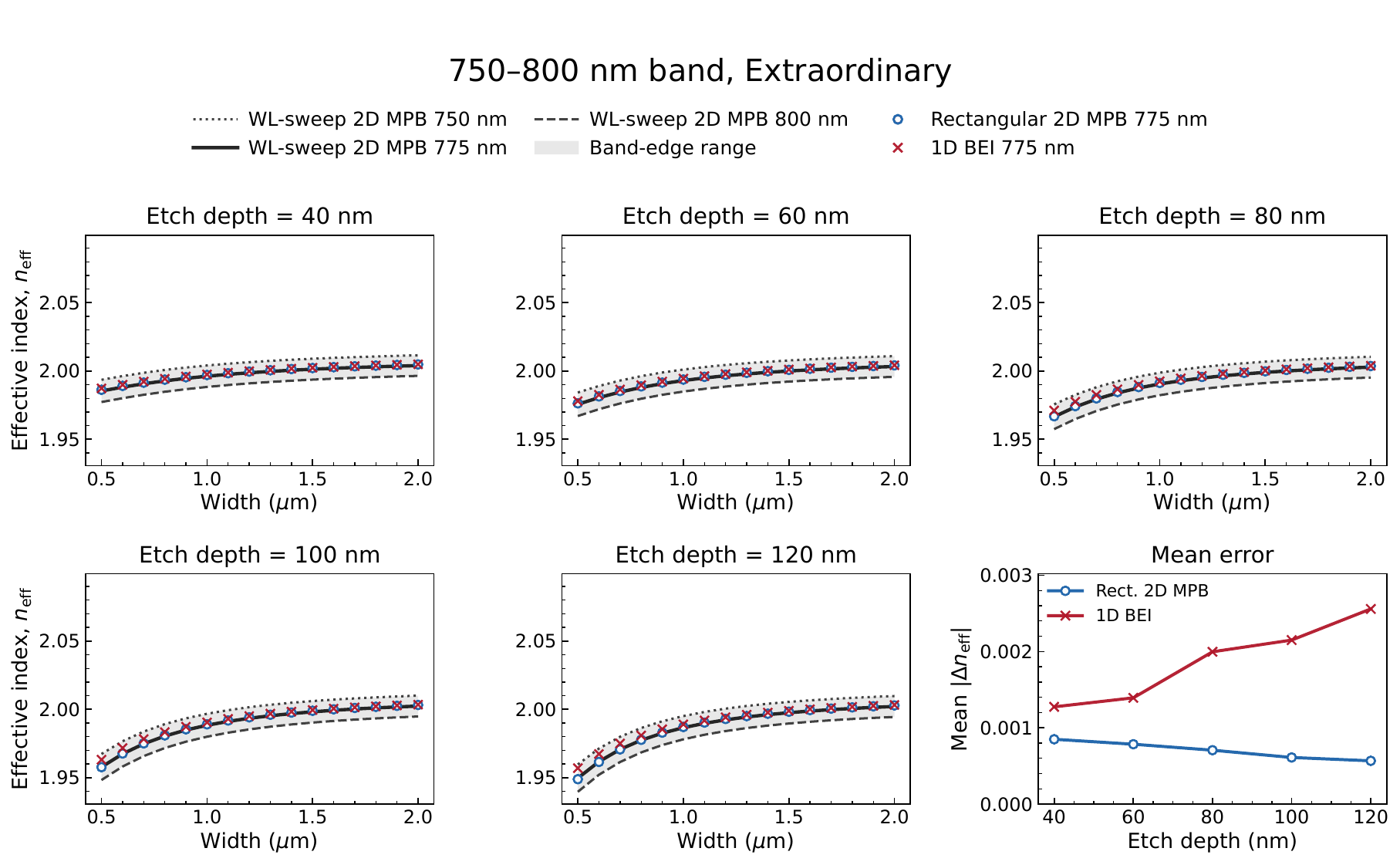}\
    \caption{%
        Comparison of the calculated effective indices at different etch depths for the extraordinary propagation direction in near-visible band.
    }
    \label{fig:S4}
\end{figure}

\begin{figure}[h!]
    \centering
    \includegraphics[width=\linewidth]{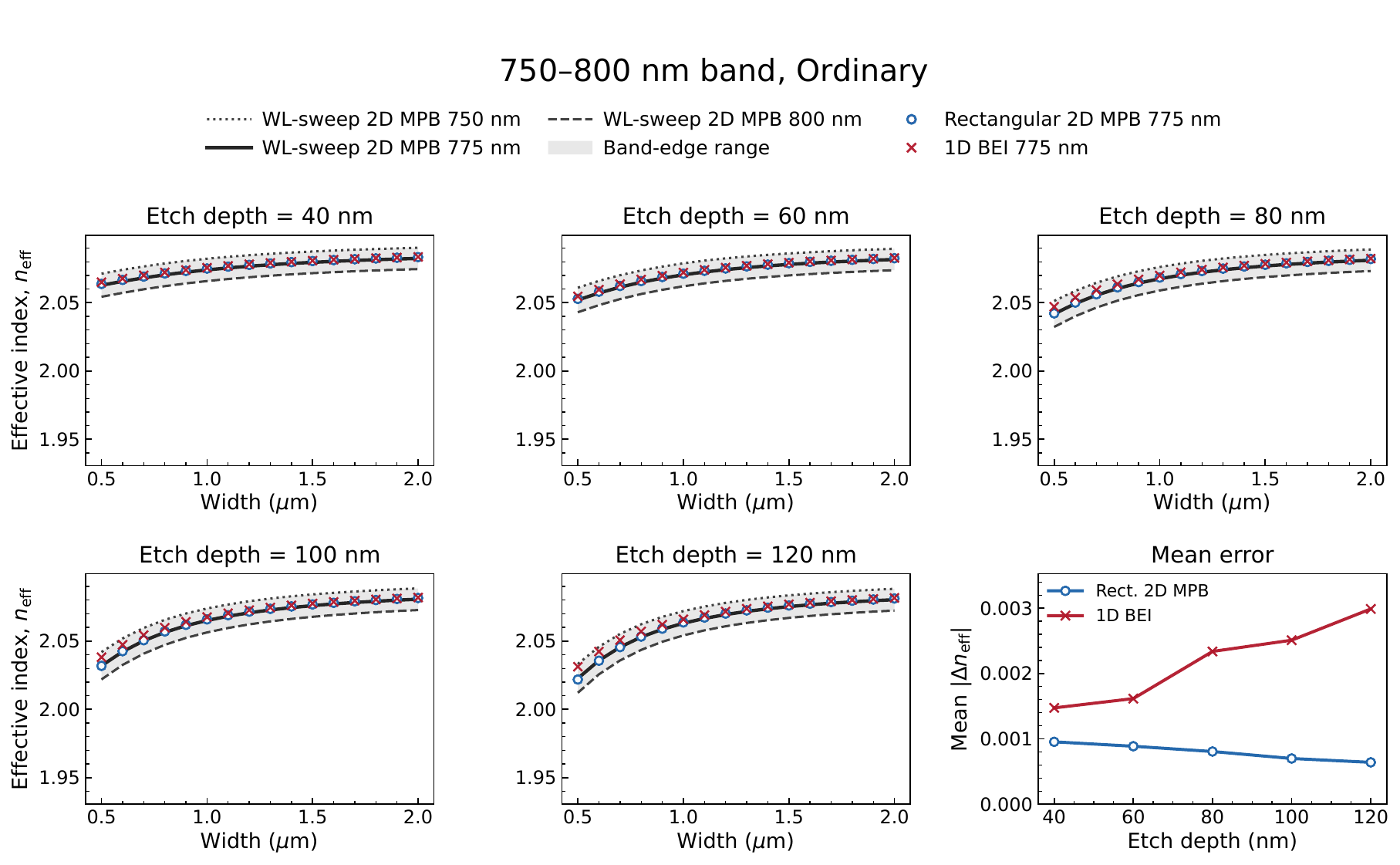}\
    \caption{%
        Comparison of the calculated effective indices at different etch depths for the ordinary propagation direction in near-visible band.
    }
    \label{fig:S5}
\end{figure}

\clearpage


\phantomsection
\label{sec:S3}
{\large\textbf{S3. Design Region Size Optimization}}

\begin{figure}[h]
    \centering
    \includegraphics[width=\linewidth]{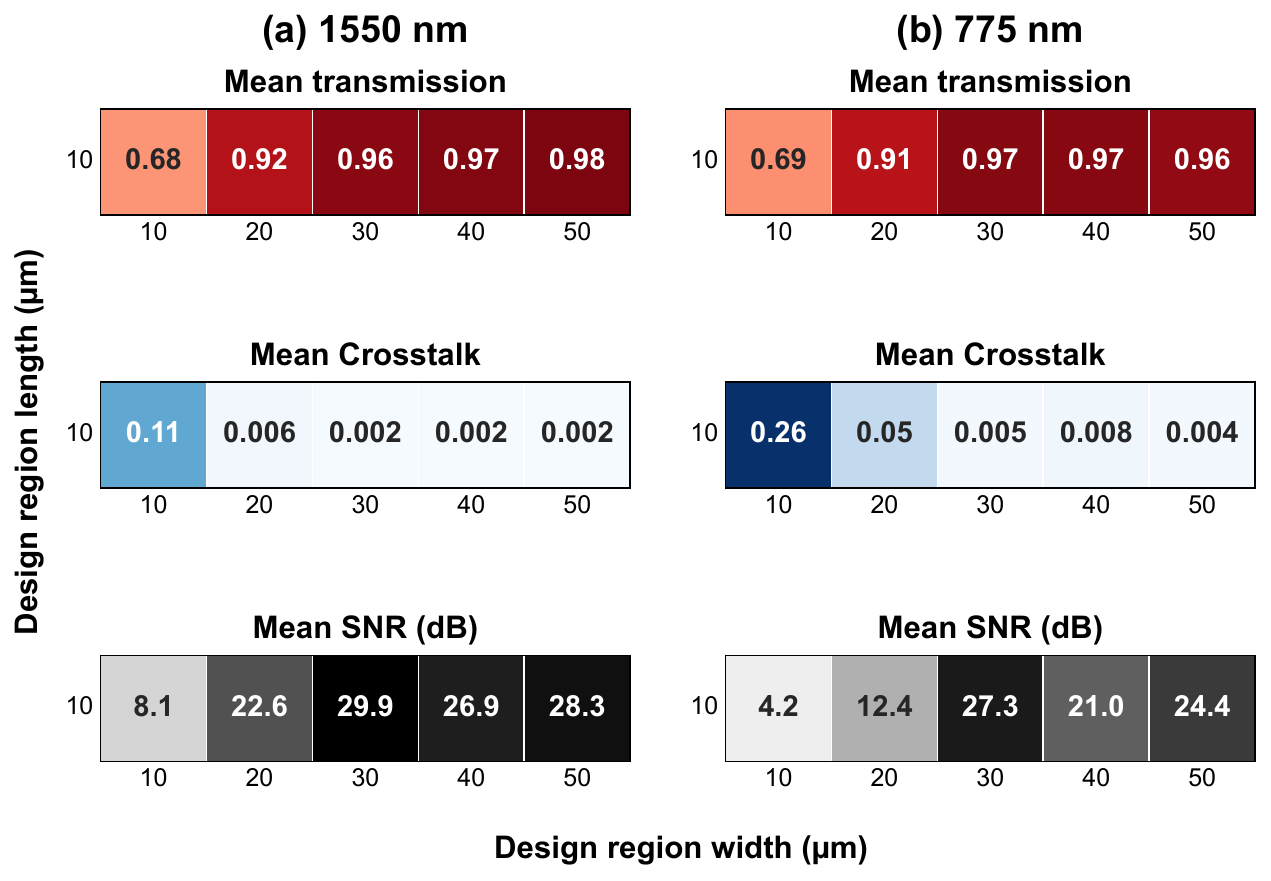}
    \caption{
    Comparison of the performance metrics obtained from the BEI-based optimization for different design-region sizes. Panels (a) and (b) show the results for the 1550 and 775 nm bands, respectively. In each panel, the upper, middle, and lower subpanels present the average transmission, average crosstalk, and average signal-to-noise ratio, respectively.}
    \label{fig:S6}
\end{figure}

\clearpage


\phantomsection
\label{sec:S4}
{\large\textbf{S4. Geometric Constraints for Slanted Sidewall}}

\begin{figure}[h!]
    \centering
    \includegraphics[width=\linewidth]{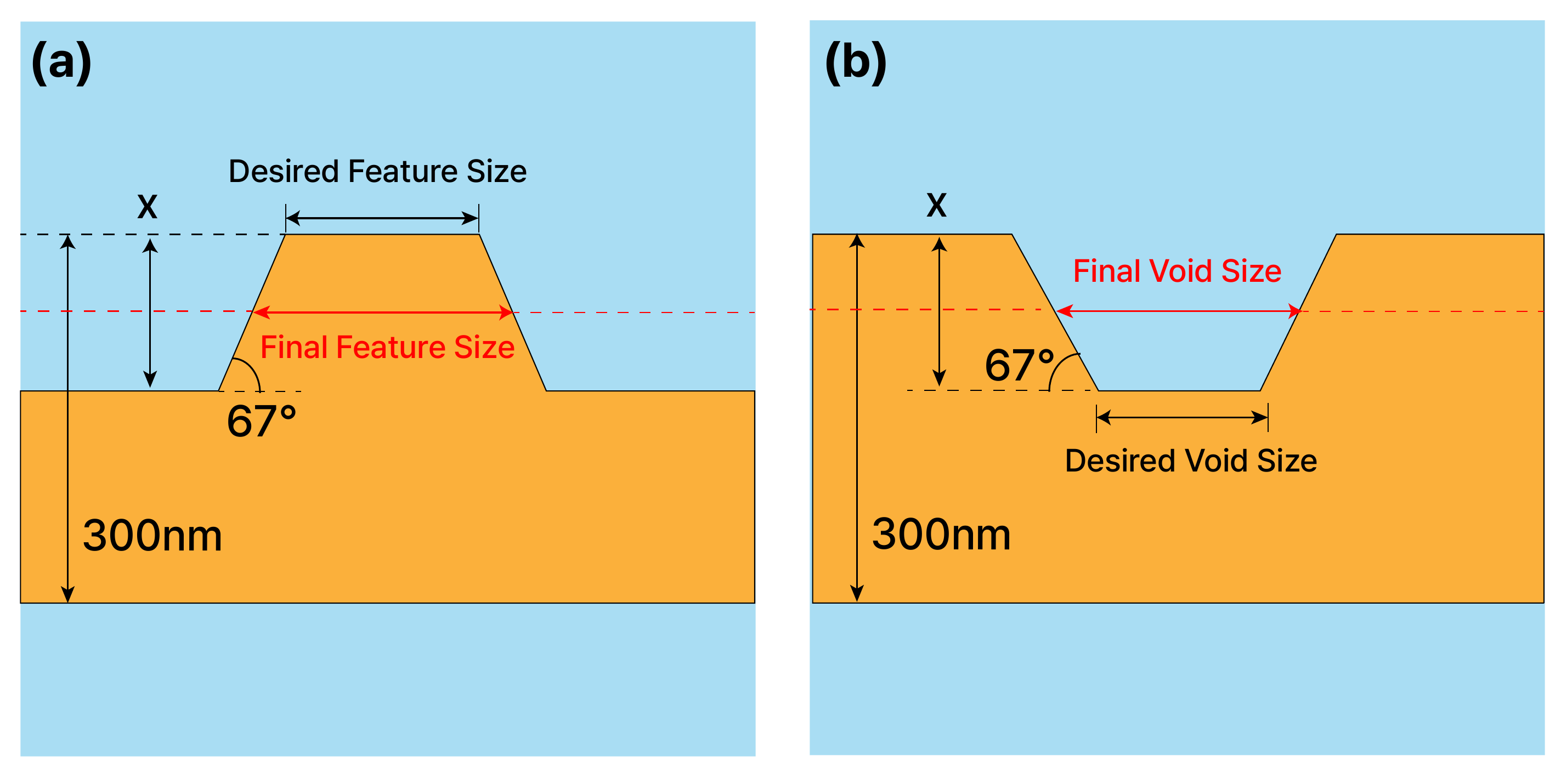}
    \caption{Schematic illustration of the determination of the geometric-constraint filter radius for structures with slanted sidewalls. (a) final feature size calculation based on the minimum feature size at the top surface. (b) final void size calculation based on the minimum void size at the substrate interface.}
    \label{fig:S7}
\end{figure}

\begin{table}[h!]
\centering
\begin{threeparttable}
\caption{Constraint values (nm) for different etch depths and desired constraint sizes. Values in parentheses indicate the discretized values according to the design region grid.}
\label{tab:constraint_values}
\begin{tabular}{c|ccccc}
\hline
\textbf{Desired constraint size} &
\multicolumn{5}{c}{\textbf{Etch depth}} \\
\cline{2-6}
 & \textbf{40 nm} & \textbf{60 nm} & \textbf{80 nm} & \textbf{100 nm} & \textbf{120 nm} \\
\hline
\textbf{50 nm}  & 50(50)  & 50(50)  & 50(50)  & 50(50)  & 50(50)  \\
\textbf{75 nm}  & 92(90)  & 100(100) & 109(110) & 117(120) & 126(130) \\
\textbf{100 nm} & 117(120) & 125(130) & 134(130) & 142(140) & 151(150) \\
\hline
\end{tabular}
\begin{tablenotes}
\footnotesize
\item For the 50 nm constraint case, no etch-depth-dependent correction was applied.
\end{tablenotes}
\end{threeparttable}
\end{table}

\newpage


\phantomsection
\label{sec:S5}
{\large\textbf{S5. Optimization Results for Different Geometric Constraints}}

\begin{figure}[h!]
    \centering
    \includegraphics[width=0.8\linewidth]{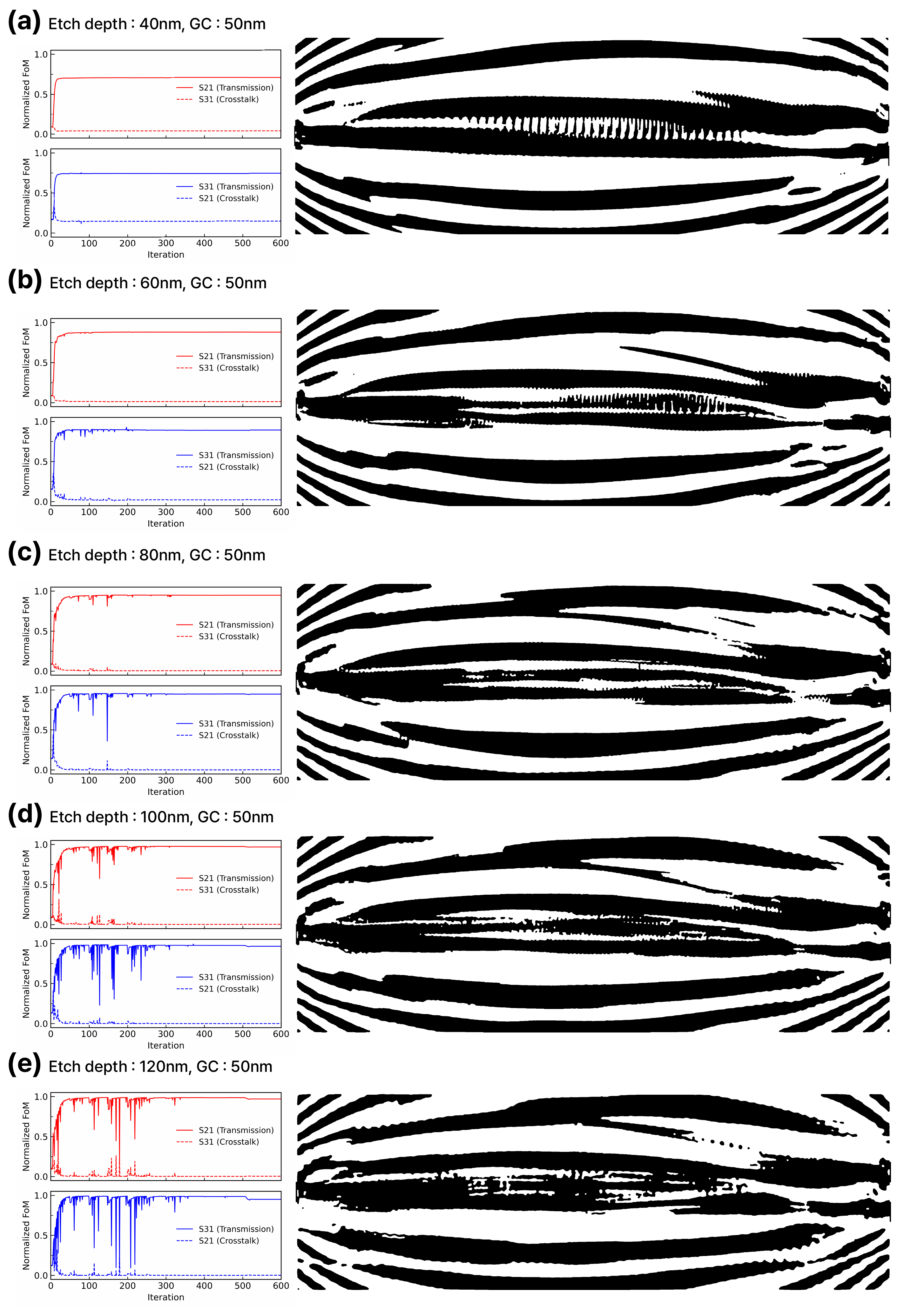}
    \caption{Optimization results at different etch depths under 50 nm geometric constraint. Panels (a)–(e) correspond to etch depths of 40, 60, 80, 100, and 120 nm, respectively. In each panel, the left plot shows the evolution of the figures of merit (FoMs) over the optimization iterations. The red and blue curves represent the telecom and near-visible bands, respectively, while the solid and dashed lines indicate transmission and crosstalk, respectively. The right side of each panel shows the corresponding optimized structure.}
    \label{fig:S8}
\end{figure}

\begin{figure}[h!]
    \centering
    \includegraphics[width=0.9\linewidth]{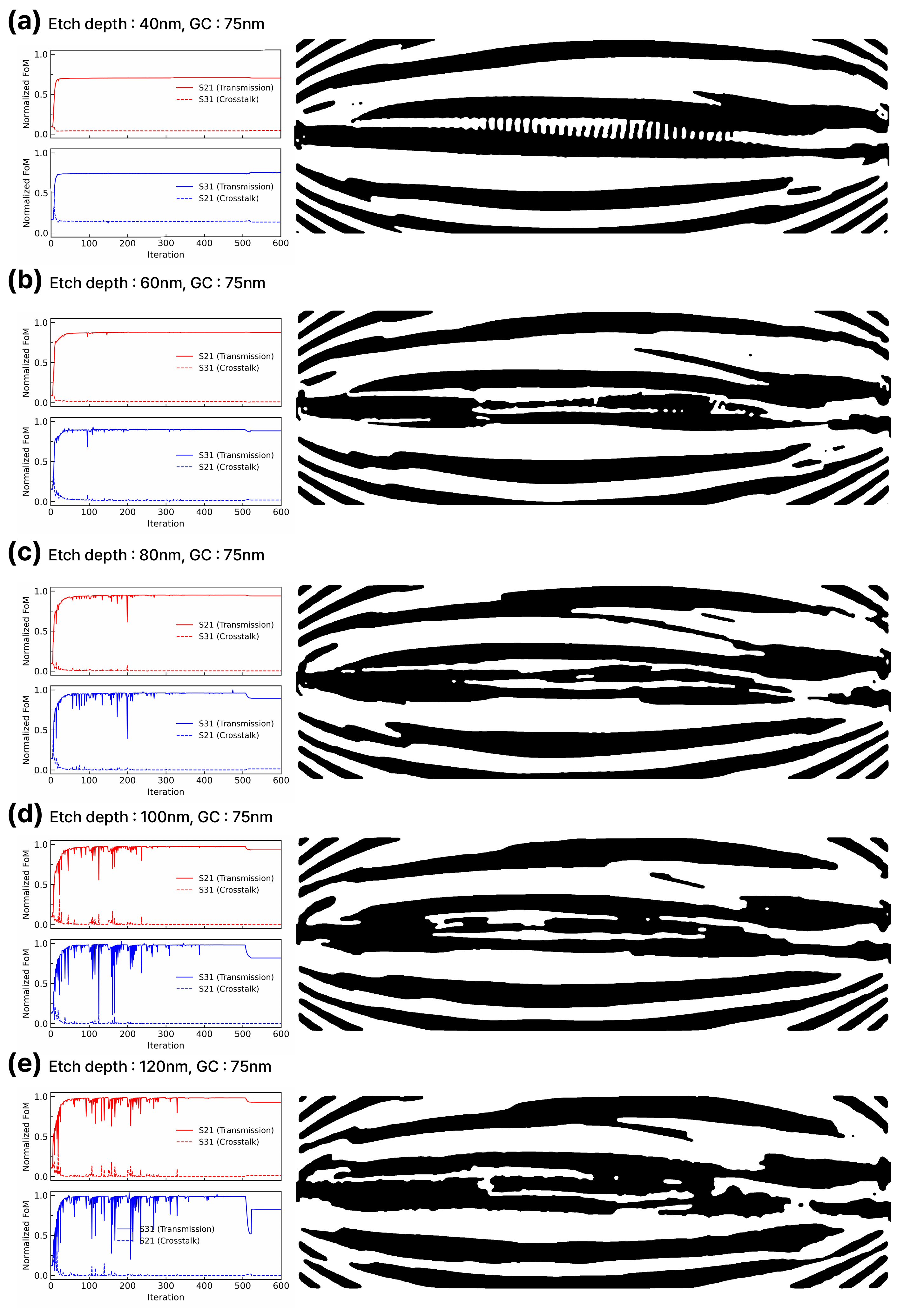}
    \caption{Optimization results at different etch depths under 75 nm geometric constraint.}
    \label{fig:S9}
\end{figure}

\begin{figure}[h!]
    \centering
    \includegraphics[width=0.9\linewidth]{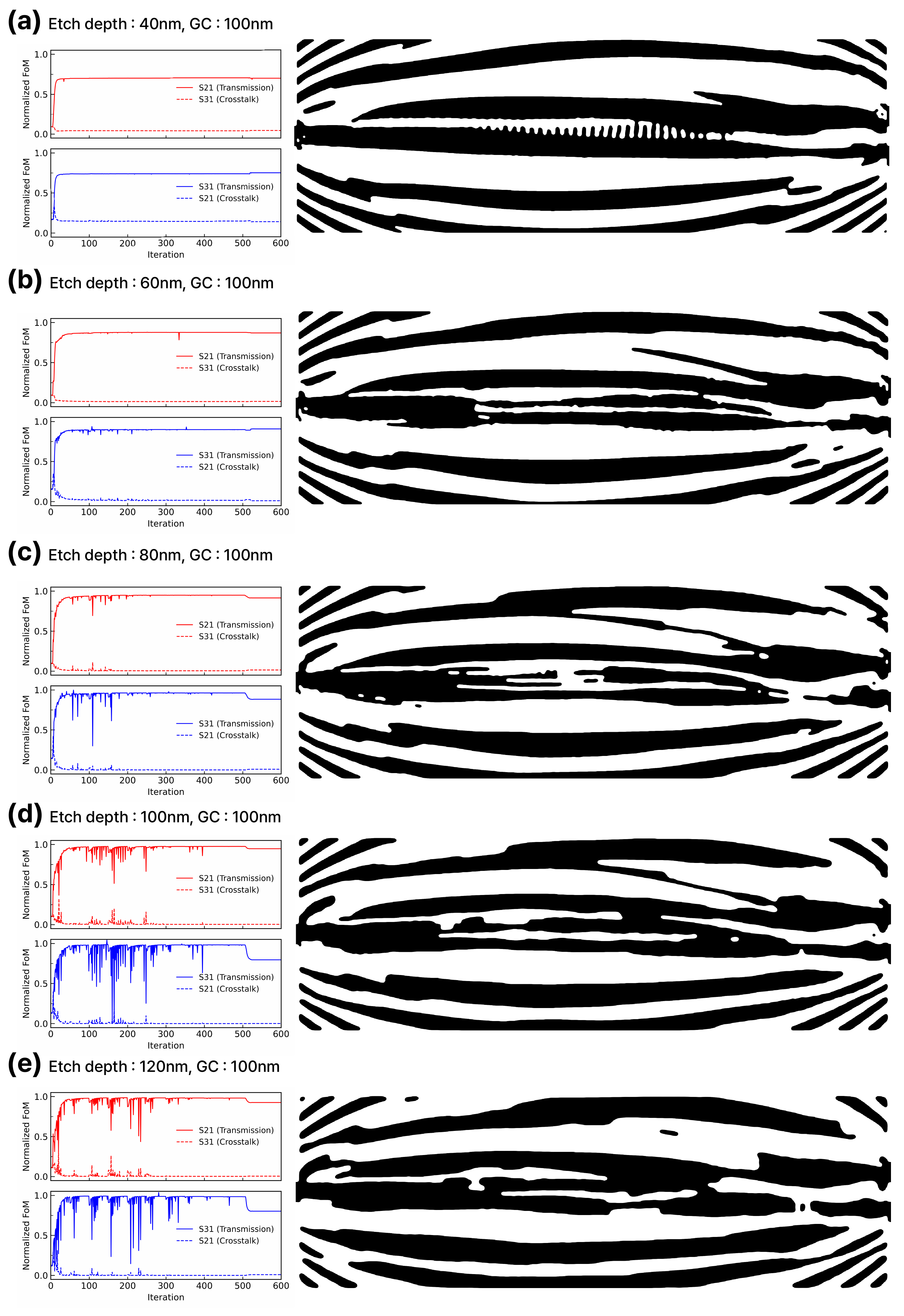}
    \caption{Optimization results at different etch depths under 100 nm geometric constraint.}
    \label{fig:S10}
\end{figure}

\clearpage


\clearpage

\phantomsection
\label{sec:S6}
{\large\textbf{S6. Optimization Result Summary}}

\begin{figure}[h!]
    \centering
    \includegraphics[width=\linewidth]{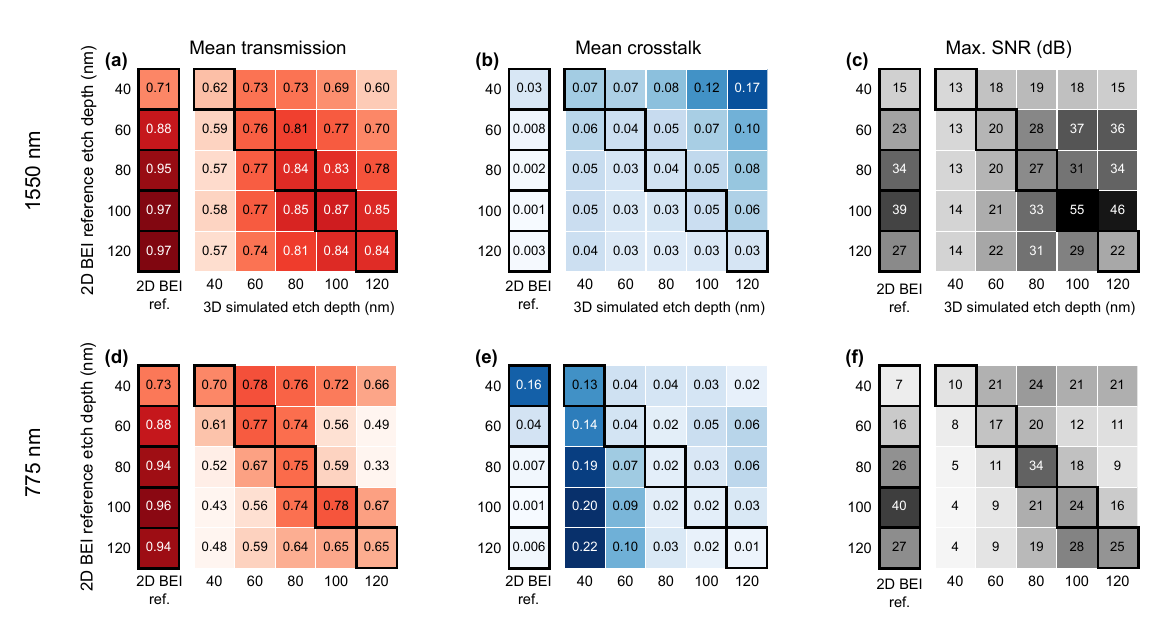}
    \caption{Comparison between the optimization results obtained using the two-dimensional BEI model and the corresponding three-dimensional verification results under a 50 nm geometric constraint. Panels (a)–(c) summarize the results for the 1550 nm band, whereas panels (d)–(f) present those for the 775 nm band. Panels (a) and (d) show the transmission to the designated output port, panels (b) and (e) show the crosstalk to the undesired output port, and panels (c) and (f) show the maximum signal-to-noise ratio. In each panel, the three-dimensional result evaluated at the reference etch depth used in the corresponding two-dimensional BEI optimization is highlighted by a black square.}
    \label{fig:S11}
\end{figure}

\begin{figure}[h!]
    \centering
    \includegraphics[width=\linewidth]{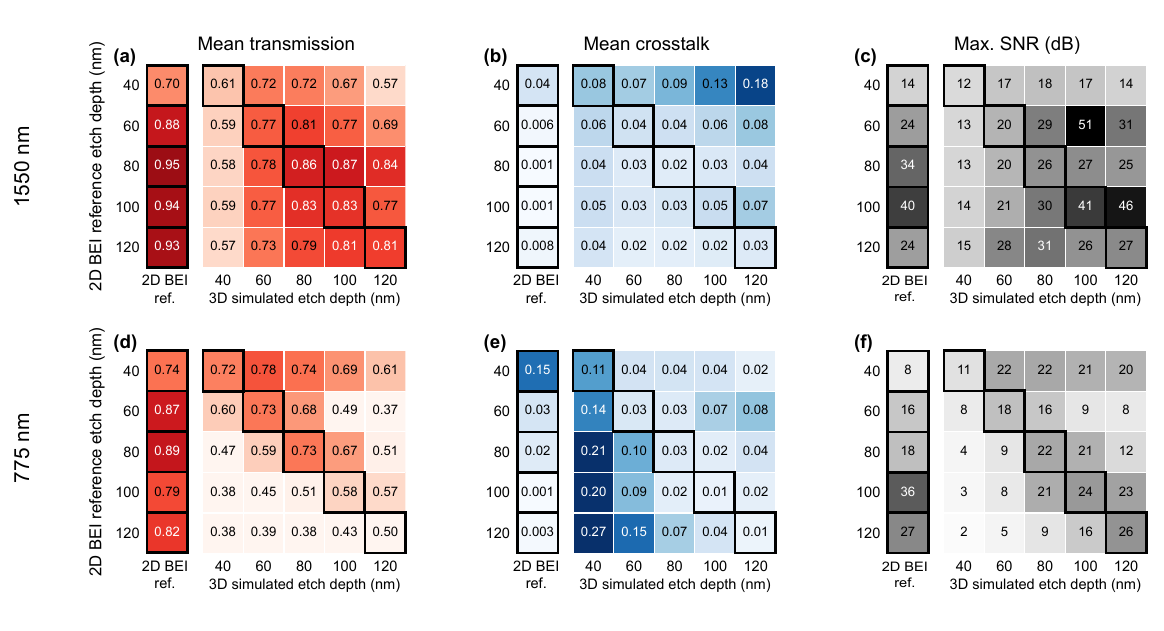}
    \caption{Comparison between the optimization results obtained using the two-dimensional BEI model and the corresponding three-dimensional verification results under 75 nm geometric constraint.}
    \label{fig:S12}
\end{figure}

\clearpage


\clearpage

\phantomsection
\label{sec:S7}
{\large\textbf{S7. Spectral Shift According to Etch Depths}}

\begin{figure}[h!]
    \centering
    \includegraphics[width=\linewidth]{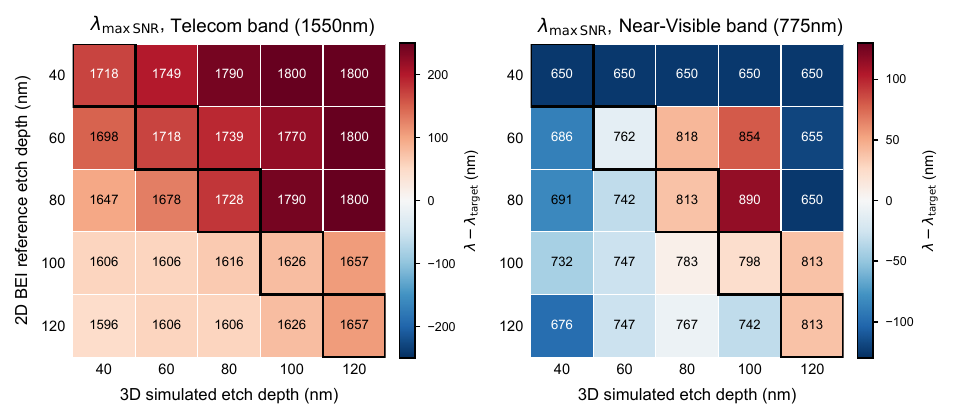}
    \caption{Wavelengths corresponding to the maximum signal-to-noise ratio obtained from the three-dimensional verification simulations. Panels in the left and right show the results for the 1550 and 775 nm bands, respectively. The vertical axis denotes the etch depth assumed in the two-dimensional BEI optimization, while the horizontal axis denotes the etch depth used in the corresponding three-dimensional simulation. The value in each cell represents the wavelength at which the signal-to-noise ratio is maximized, and the color indicates its deviation from the target wavelength. Black outlines highlight the matched cases in which the two-dimensional reference etch depth and the three-dimensional simulated etch depth are identical.}
    \label{fig:S13}
\end{figure}

\begin{figure}[h!]
    \centering
    \includegraphics[width=\linewidth]{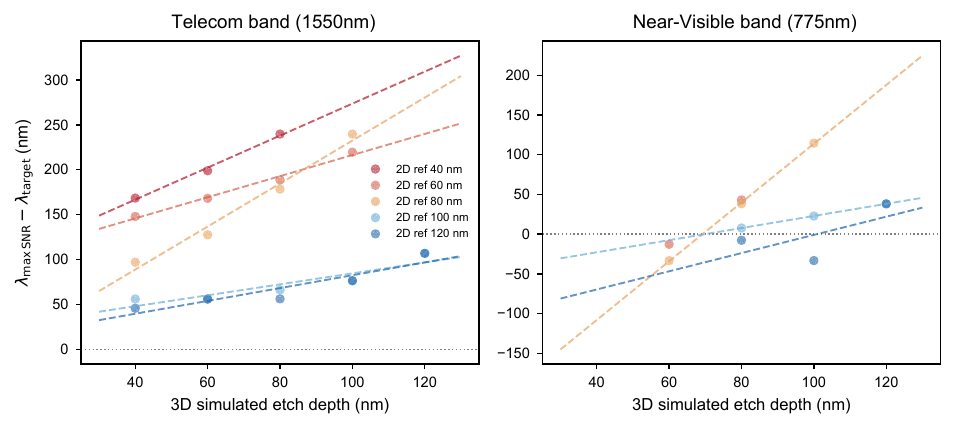}
    \caption{Shift in the wavelength corresponding to the maximum signal-to-noise ratio as a function of the three-dimensional simulated etch depth. Panels (a) and (b) show the results for the 1550 and 775 nm bands, respectively. Each set of data points corresponds to a structure optimized using the two-dimensional BEI model at a fixed reference etch depth, while the three-dimensional etch depth is varied. Dashed lines indicate linear fits to the valid data points, excluding cases in which the maximum-SNR wavelength was located near the simulated spectral boundary or the maximum signal-to-noise ratio was below 10 dB. The horizontal dotted line indicates zero wavelength shift relative to the target wavelength.}
    \label{fig:S13}
\end{figure}

\clearpage


\phantomsection
\label{sec:S8}
{\large\textbf{S8. SEM Images}}

\begin{figure}[h!]
    \centering
    \includegraphics[width=70mm]{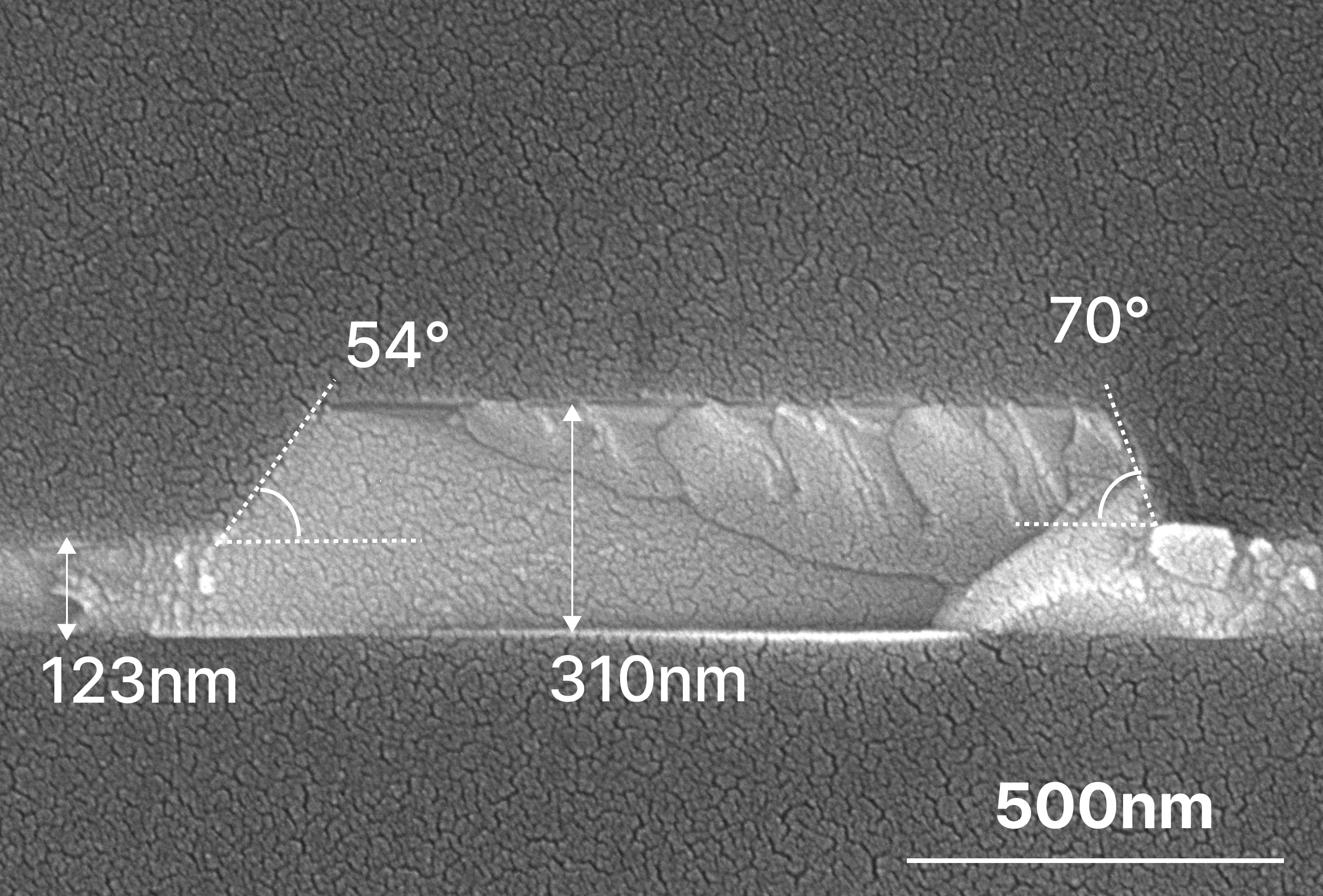}
    \caption{Cross-sectional scanning electron microscopy (SEM) image of the fabricated waveguide}
    \label{fig:S15}
\end{figure}

\begin{figure}[h!]
    \centering
    \includegraphics[width=\linewidth]{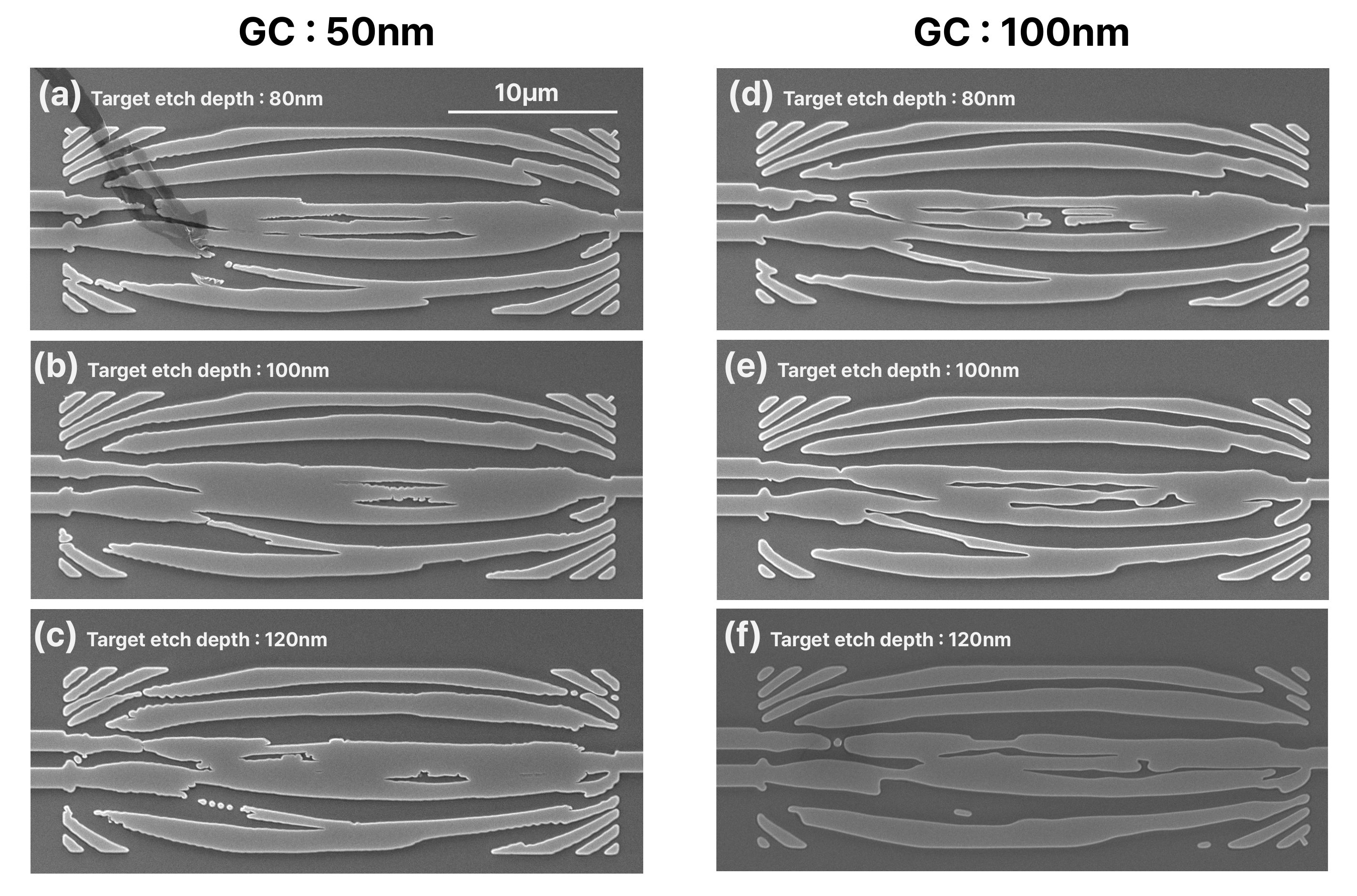}
    \caption{Top-view scanning electron microscopy (SEM) images of the fabricated LN wavelength demultiplexers. Panels (a)–(c) show devices designed with a 50 nm geometric constraint, whereas panels (d)–(f) show devices designed with a 100 nm geometric constraint. The target etch depths are 80 nm for panels (a) and (d), 100 nm for panels (b) and (e), and 120 nm for panels (c) and (f).}
    \label{fig:S16}
\end{figure}

\clearpage


\phantomsection
\label{sec:S9}
{\large\textbf{S9. Fabrication Methods}}

\begin{figure}[h!]
    \centering
    \includegraphics[width=\linewidth]{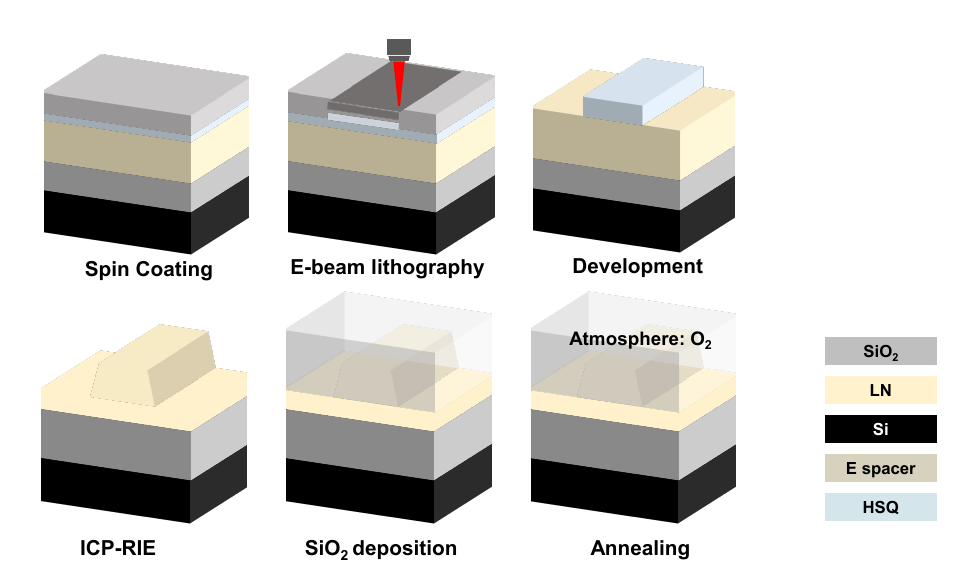}
    \caption{Schematic illustration of the fabrication process.}
    \label{fig:S17}
\end{figure}

\clearpage


\phantomsection
\label{sec:S10}
{\large\textbf{S10. Measurement Setup}}

\begin{figure}[h!]
    \centering
    \includegraphics[width=\linewidth]{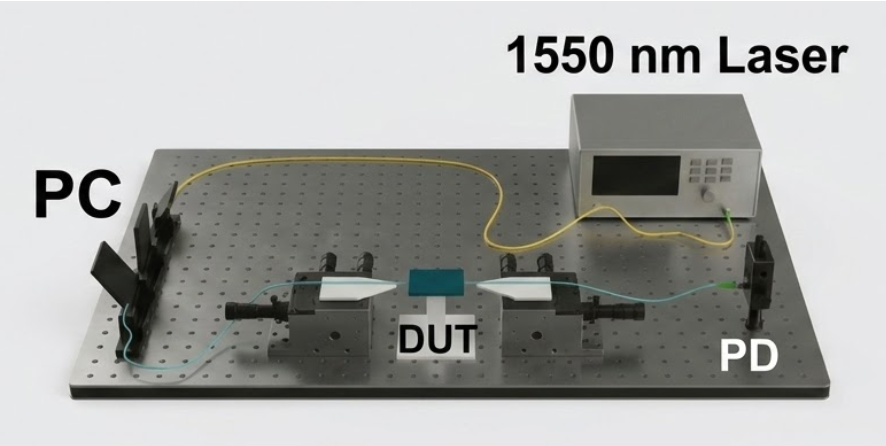}
    \caption{Schematic illustration of measurement setup.}
    \label{fig:S18}
\end{figure}

\clearpage


\phantomsection
\label{sec:S11}
{\large\textbf{S11. Measurement Result}}

\begin{figure}[h!]
    \centering
    \includegraphics[width=\linewidth]{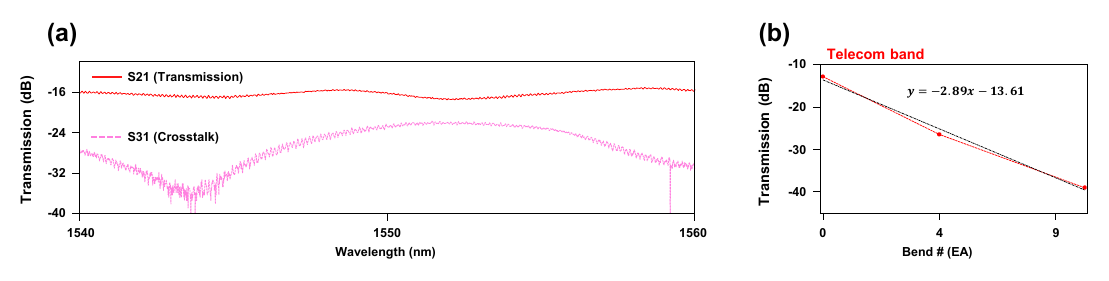}
    \caption{Measurement results for the device designed with a target etch depth of 120 nm and fabricated using a 120 nm etch. (a) Measured transmission and crosstalk spectra in the telecom band, represented by a red solid line and a pink dashed line, respectively. (b) Cumulative bending loss as a function of the cascaded WDM devices. The shallow etch depth of the waveguide structure weakens the optical mode confinement, causing the light to radiate heavily at the routing curves. This structurally induced bending loss becomes the dominant limiting factor for the overall transmission efficiency.}
    \label{fig:S19}
\end{figure}

\begin{figure}[h!]
    \centering
    \includegraphics[width=\linewidth]{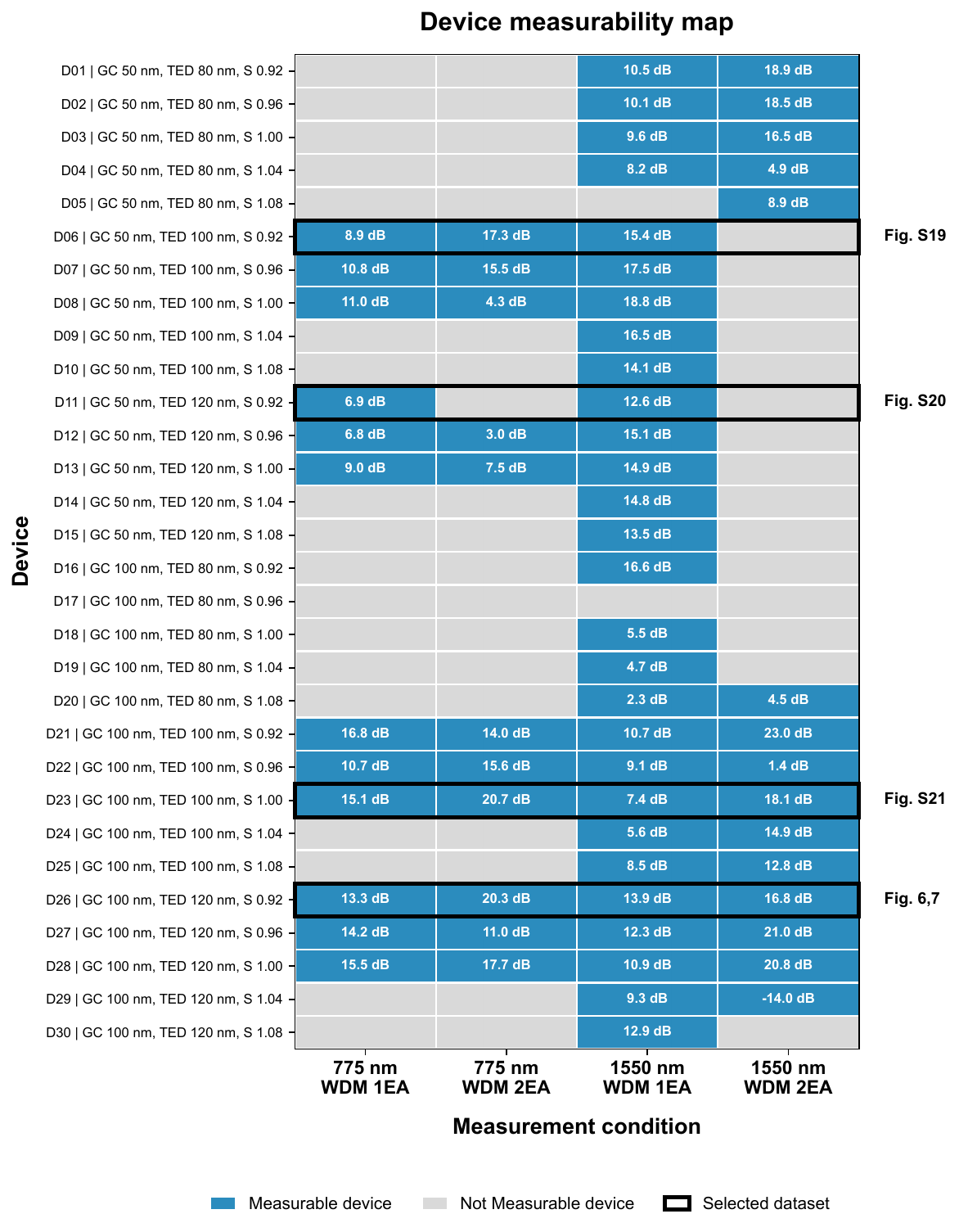}
    \caption{Device measurability map for the fabricated designs with etch depth of 180nm. The device set comprises 30 design conditions defined by two geometric constraints (50 and 100 nm), three target etch depths (80, 100, and 120 nm), and five scaling factors ranging from 0.92 to 1.08. The measurability of configurations containing one and two cascaded wavelength demultiplexers is compared for each design condition. Blue cells indicate measurable devices, whereas gray cells indicate devices that could not be measured because of fabrication-related defects, such as contamination or disconnected waveguides. Representative results presented in the main text and Supporting Information are highlighted by black squares.}
    \label{fig:S20}
\end{figure}

\begin{figure}[h!]
    \centering
    \includegraphics[width=\linewidth]{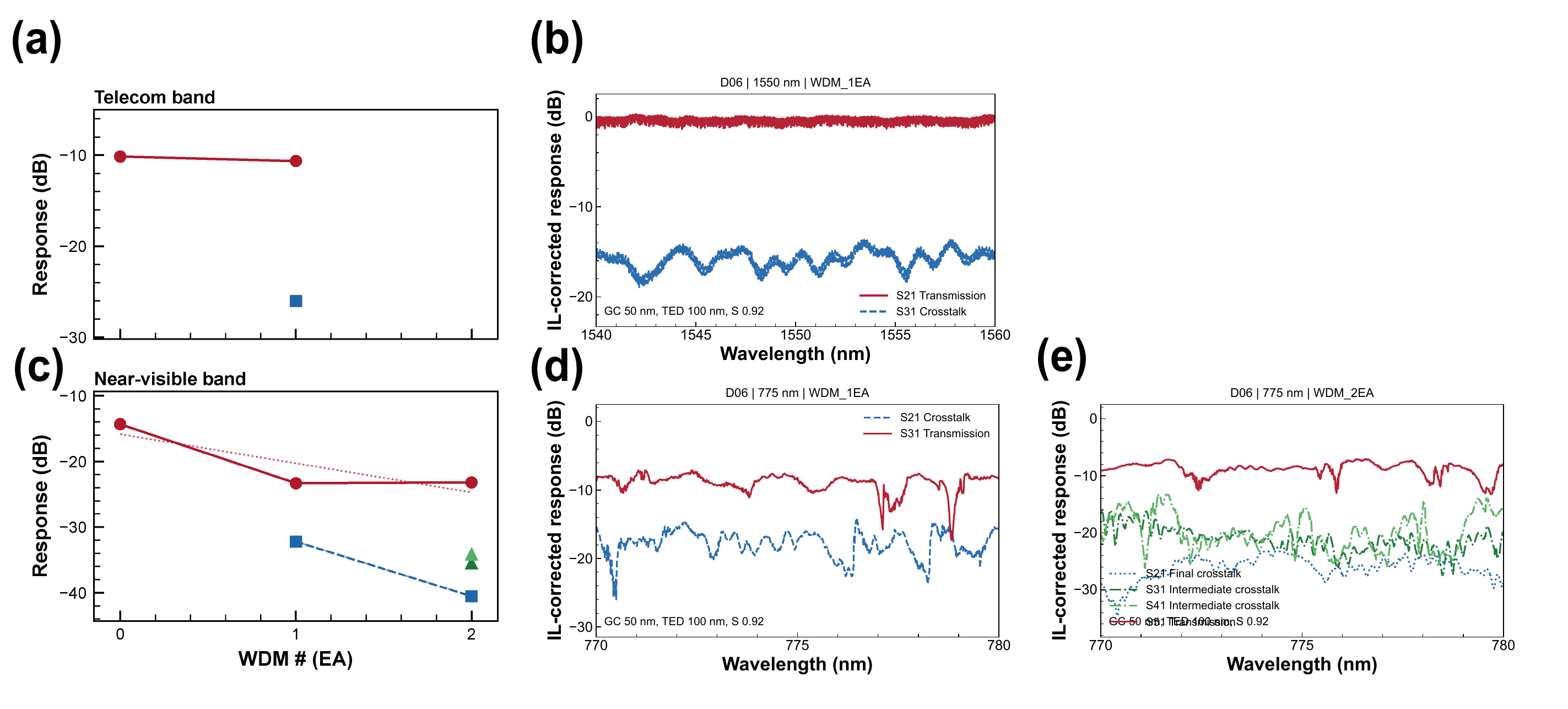}
    \caption{Measured responses of device 6, designed with a 50 nm geometric constraint, a target etch depth of 100 nm, and a scaling factor of 0.92. Panels (a) and (b) show the telecom-band results, whereas panels (c)–(e) show the near-visible-band results. Panels (a) and (c) present the cumulative insertion loss as a function of the number of cascaded WDM devices. Panels (b) and (d) show the insertion-loss-corrected responses for a single WDM device, while panel (e) shows the insertion-loss-corrected response for two cascaded WDM devices.}
    \label{fig:S21}
\end{figure}

\begin{figure}[h!]
    \centering
    \includegraphics[width=0.6\linewidth]{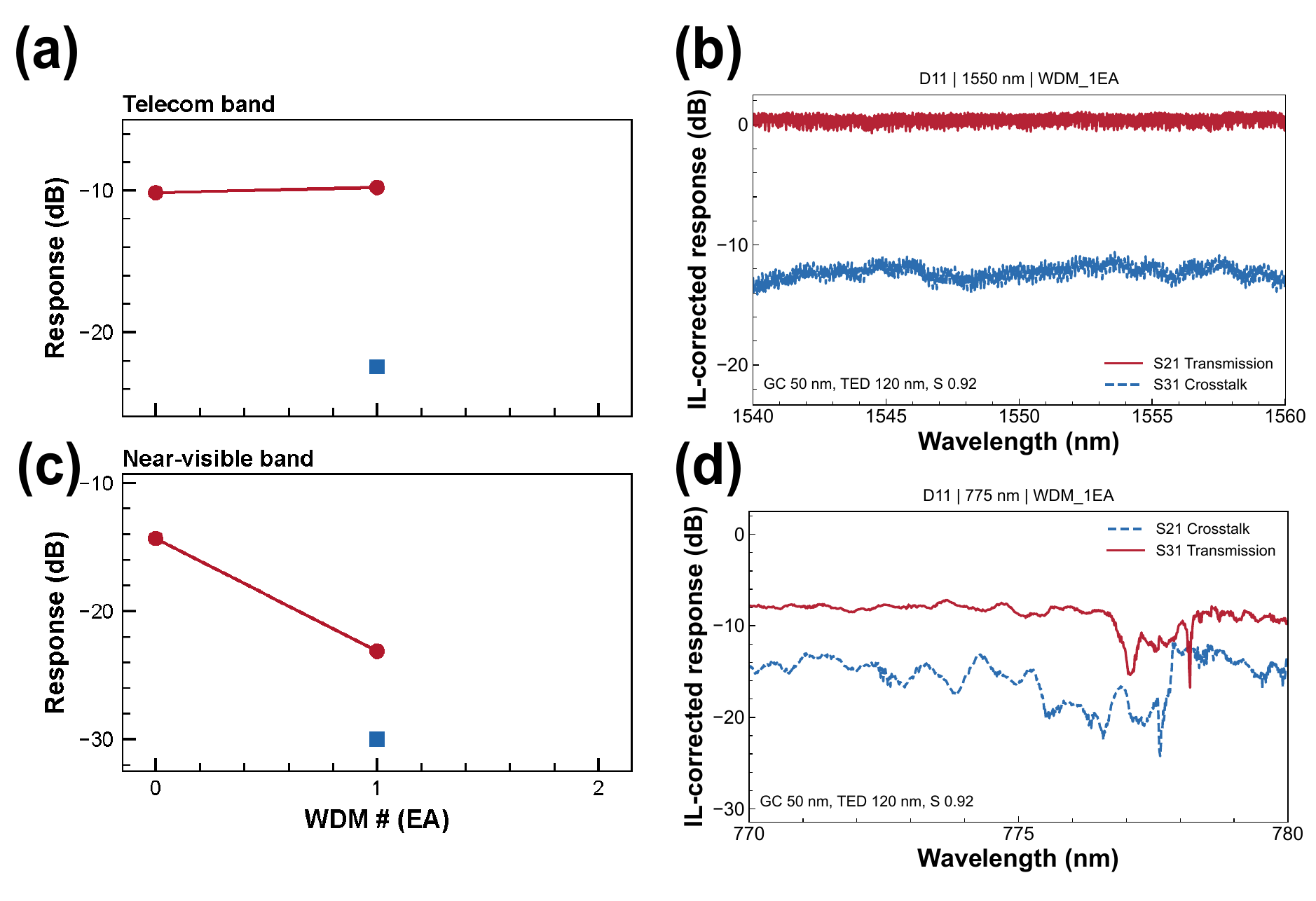}
    \caption{Measured responses of device 11, designed with a 50 nm geometric constraint, a target etch depth of 120 nm, and a scaling factor of 0.92. Panels (a) and (b) show the telecom-band results, whereas panels (c) and (d) show the near-visible-band results. Panels (a) and (c) present the cumulative insertion loss as a function of the number of cascaded WDM devices, while panels (b) and (d) show the measured responses of a single WDM device.}
    \label{fig:S22}
\end{figure}

\begin{figure}[h!]
    \centering
    \includegraphics[width=\linewidth]{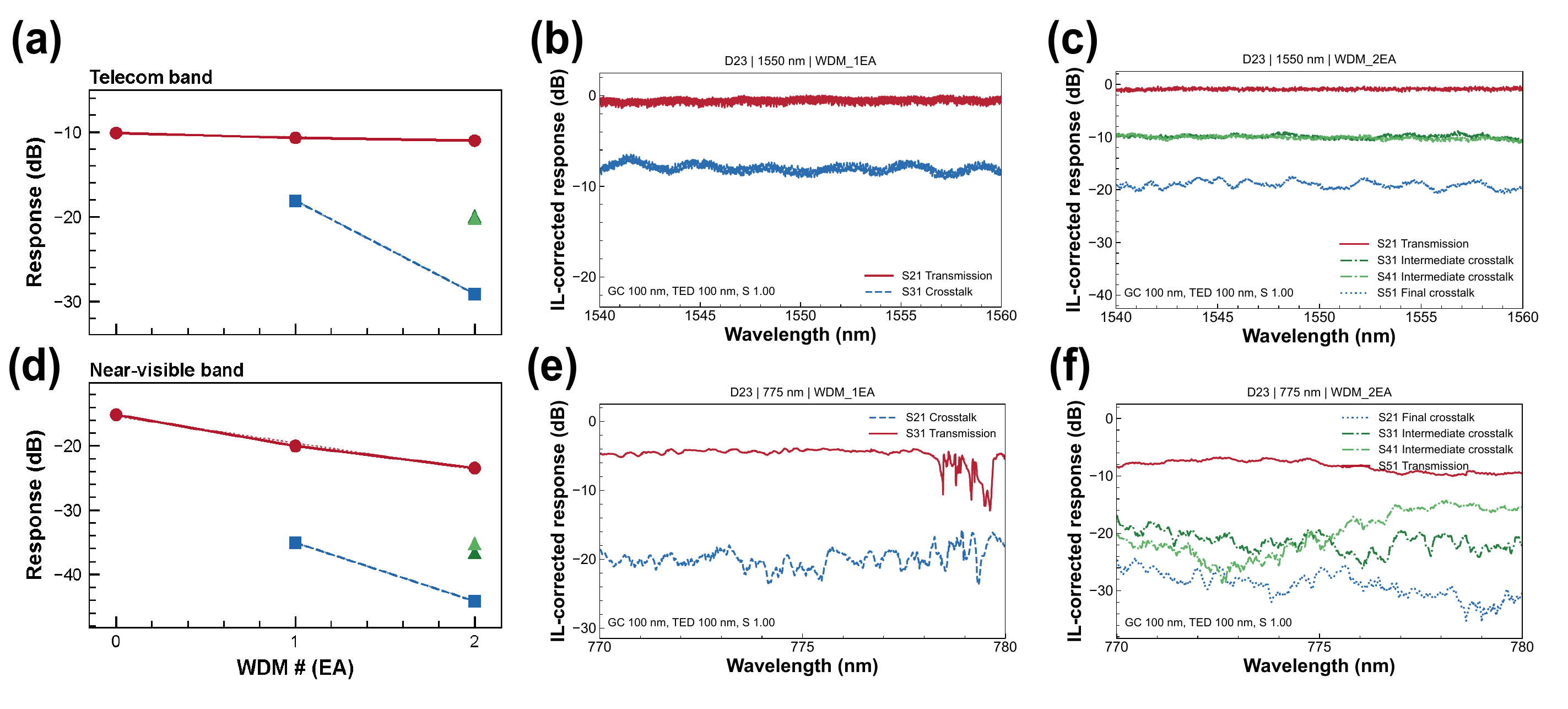}
    \caption{Measured responses of device 23, designed with a 100 nm geometric constraint, a target etch depth of 100 nm, and a scaling factor of 1.00. Panels (a)–(c) show the telecom-band results, whereas panels (d)–(f) show the near-visible-band results. Panels (a) and (d) present the cumulative insertion loss as a function of the number of cascaded WDM devices. Panels (b) and (e) show the measured responses of a single WDM device, while panels (c) and (f) show the measured responses of two cascaded WDM devices.}
    \label{fig:S23}
\end{figure}

\clearpage